\documentclass[acmsmall]{acmart}
\renewcommand\footnotetextcopyrightpermission[1]{} 
\setcopyright{none}

\usepackage{booktabs}   
\usepackage{subcaption} 

\usepackage{float}
\usepackage[english]{babel}

\usepackage[T1]{fontenc}
\usepackage[scaled=0.85]{beramono}

\usepackage{todonotes}

\usepackage{framed}
\usepackage{hyperref}
\usepackage{xspace}
\usepackage{fancyvrb}
\usepackage{listings}
\usepackage{booktabs}
\hypersetup{colorlinks=true,allcolors=blue}
\usepackage{array}
\usepackage{url}
\usepackage{comment}
\usepackage[inline]{enumitem}
\usepackage{pifont}

\usepackage{stmaryrd}
\usepackage{xcolor}
\usepackage{mathtools}
\usepackage{xspace}
\usepackage{mathpartir}
\usepackage{graphicx}
\usepackage{caption}
\usepackage{xcolor}
\usepackage{tikz}
\usetikzlibrary{arrows.meta,automata,shapes,patterns}
\usepackage{listings}

\ifdefined\withcolor
  \definecolor{haskellstr}{rgb}{0.2, 0.2, 0.6}
  \definecolor{haskellred}{rgb}{1.0, 0.0, 0.0}
  \definecolor{gray_ulisses}{gray}{0.55}
  \definecolor{green_ulises}{rgb}{0.2,0.75,0}
  \definecolor{haskelltypes}{rgb}{0.71,0.33,0.14}
  \definecolor{logiccolor}{rgb}{0.0, 0.0, 1.0}
\else
  \definecolor{haskellstr}{rgb}{0.2, 0.2, 0.6}
  \definecolor{haskellred}{gray}{0.1}
  \definecolor{gray_ulisses}{gray}{0.55}
  \definecolor{green_ulisses}{gray}{0.1}
  \definecolor{haskelltypes}{gray}{0.1}
  \definecolor{logiccolor}{gray}{0.1}
\fi

\definecolor{haskellblue}{rgb}{0.0, 0.0, 1.0}
\definecolor{haskell_green}{rgb}{0.0, 0.5, 0.0}
\definecolor{blue_violet}{rgb}{0.54, 0.17, 0.89}
\definecolor{castanho_ulisses}{rgb}{0.43, 0.21, 0.1}
\definecolor{preto_ulisses}{rgb}{0.21,0.00,0.80}

\definecolor{subtleOpHighlight}{rgb}{0.4, 0.2, 0.0}

\definecolor{lcolor}{gray}{0.0}
\definecolor{lappcolor}{gray}{0.0}
\definecolor{lappascolor}{gray}{0.0}

\def\codesize{\normalsize}
\def\incodesize{\normalsize}

\newcommand\showfocus[1]{\color{purple}{\textbf{#1}}}
\newcommand\showlogic[1]{\color{logiccolor}{#1}}

\lstdefinelanguage{HaskellUlisses} {
  aboveskip=\smallskipamount,
  belowskip=\smallskipamount,
  basicstyle=\ttfamily\codesize,
  moredelim=[is][\showfocus]{\#}{\#},
  moredelim=[is][\showlogic]{!}{!},
  sensitive=true,
  morecomment=[l][\color{gray_ulisses}\ttfamily\itshape\codesize]{--},
  morestring=[b]",
  stringstyle=\color{haskellstr},
  showstringspaces=false,
  numberstyle=\codesize,
  numberblanklines=true,
  showspaces=false,
  breaklines=true,
  showtabs=false,
  literate={
           {<!}{{{\color{lcolor}<!}}}2
           {`}{{{$^{\backprime}{}$}}}1
           {?}{{{\color{lcolor}?}}}1
           {<=}{{$\leq$}}1
           {/=}{{$\neq$}}1
           {bot}{{$\bot$}}1
           {top}{{$\top$}}1
           {theta}{{$\theta$}}1
           {gf}{{{\color{lappascolor}f}}}1
           {rmap}{{{\color{lappcolor}map}}}3
           {gmap}{{{\color{lappascolor}map}}}3
           {r.}{{{\color{lappcolor}.}}}1
           {g.}{{{\color{lappascolor}.}}}1
           {r++}{{{\showfocus{++}}}}2
           {g++}{{{\color{lappascolor}++}}}2
           {>>}{{{\color{haskellblue}>>}}}2
           {>>=}{{{\color{haskellblue}>>=}}}3
           {</>}{{{\color{haskellblue}</>}}}3
           {<*>}{{{\color{haskellblue}<*>}}}3
           {++}{{{\color{haskellblue}++}}}3
           {gfib}{{{\color{lappascolor}fib}}}3
           {rfib}{{{\color{lappcolor}fib}}}3
           {r++}{{{\color{lappcolor}++}}}2
           {env}{{$\Gamma$}}1
           {|-}{{$\vdash\;$}}1
           {<=!}{{{\color{lcolor}<=!}}}3
           {!=}{{$\neq$}}1
           {~>}{{$\imparrow$}}2
           {<-}{{$\leftarrow$}}2
           {->}{{$\rightarrow$}}2                    
           {:=}{{$\defeq$}}2                          
           {Gamma}{{$\Gamma$}}1                        
           {dollar}{{$\texttt{\$}$}}1
           {Set_mem}{{$\in$}}1
           {Set_cup}{{$\cup$}}1
           {Set_cap}{{$\cap$}}1
           {Set_emp}{{$\emptyset$}}1
           {Set_sub}{{$\subseteq$}}1
           {<=>}{{$\Leftrightarrow$}}3
           {=>}{{$\Rightarrow$}}2
           {1->}{{$\rightarrow$}}1
           {1=>}{{$\Rightarrow$}}1
           {||-}{{$\vdash$}}1
           {|->}{{$\mapsto$}}2
           {<:}{{$\preceq$}}1
           {Inarritu}{Inarritu}8},
  emph=
  {[1]
    succ,incr,two,incrMany,three,id,get,set,return,grant,revoke,main,canRead,safeRead,
    test1,test2,
    gnt,rev,next,foo,bar,baz,pure,client,done,clynt,fin,
    FilePath,IOError,abs,acos,acosh,all,and,any,appendFile,approxRational,asTypeOf,asin,
    asinh,atan,atan2,atanh,basicIORun,break,catch,ceiling,chr,compare,concat,concatMap,
    cos,cosh,curry,cycle,decodeFloat,denominator,digitToInt,div,divMod,drop,
    dropWhile,either,elem,encodeFloat,enumFrom,enumFromThen,enumFromThenTo,enumFromTo,
    error,even,exponent,fail,mapMaybe,filter,flip,floatDigits,floatRadix,floatRange,floor,
    foldl,foldl1,foldr1,fromDouble,fromEnum,fromInt,fromInteger,fromIntegral,
    fromRational,fst,gcd,put,tick,tock,tocker,ticker,getChar,getContents,getLine,head,inRange,index,init,intToDigit,
    interact,ioError,isAlpha,isAlphaNum,isAscii,isControl,isDenormalized,isDigit,isHexDigit,
    isIEEE,isInfinite,isLower,isNaN,isNegativeZero,isOctDigit,isPrint,isSpace,isUpper,iterate,
    last,lcm,length,lex,lexDigits,lexLitChar,lines,log,logBase,lookup,mapM,mapM_,max,
    maxBound,posMax,negMax,maximum,maybe,min,minBound,minimum,mod,negate,notElem,null,numerator,odd,
    or,ord,pi,pred,primExitWith,print,product,putChar,putStr,putStrLn,quot,
    quotRem,range,rangeSize,read,readDec,readFile,readFloat,readHex,readIO,readInt,readList,readLitChar,
    readLn,readOct,readParen,readSigned,reads,readsPrec,realToFrac,recip,rem,repeat,replicate,return,
    reverse,round,scaleFloat,scanl,scanl1,scanr,scanr1,seq,sequence,sequence_,show,showChar,showInt,
    showList,showLitChar,showParen,showSigned,showString,shows,showsPrec,significand,signum,sin,
    sinh,snd,span,splitAt,sqrt,subtract,tail,take,takeWhile,tan,tanh,threadToIOResult,toEnum,
    toInteger,toLower,toRational,toUpper,truncate,uncurry,undefined,unlines,until,unwords,unzip,
    unzip3,userError,words,writeFile,zip,zip3,zipWith,zipWith3,listArray,doParse,empty,for,initTo,
        assert,compose,checkGE,maxEvens,empty,create,get,set,initialize,idVec,fastFib,fibMemo,
        ex1,ex2,ex3,inc,dec,isPos,positives,find,flatten, expand,exAll,
        ind,evenLen,lenAppend,exDistOr,allDistAnd,len,size,union,singleton,initUpto,trim,
        insertSort,decsort,qsort,reverse,append,upperCase, ifM, whileM, get, decrM, diff,
        project, select, sel, elts, keys, dkeys, dfun, addKey, pTrue, emptyRD, rFalse,
                dom, rng, isI, isD, isS, movie1, movie2,  toI, toS, toD, good_titles, runState, ret,
                update, getCtr, setCtr, ctr, rdCtr, wrCtr, ifTest, whileTest, posCtr, zeroCtr, decr, decCtr,
                pread , pwrite , plookup , pcontents, pcreateF , pcreateFP, pcreateD, active, caps, pset, eqP,
                write, contents, alloc, derivP, copyP, createDir, store, copyRec, copySpec,
                preservation, progress, soundness, impossible, lemValStep,
                forM_, when, flookup, fread, createDir, pcreateFile, isFile, copyFrame, ?
  },
  emphstyle={[1]\color{haskellblue}},
  emph=
  {[2]Show,Eq,Iso,VerifiedOrd,Ord,Num,UpClosed,Comp,Wit,Witness,Inductive,Meet,Flip,TRUE,
      Peano,Nat,Pos,SInt,Neg,IntGE,Plus,List,PAnd, POr, POrL, POrR,
        Bool,Char,Double,Either,Float,IO,Integer,Int,Maybe,Up,Mono,Identity,
        Ordering,Rational,Ratio,ReadS,ShowS,File,Token,ST,String,Str,Word8,
        InPacket,Tree,Vec,NullTerm,IncrList,DecrList,
        UniqList,BST,MinHeap,MaxHeap,World,RIO,IO,HIO,Post,Pre, OptEq,
        Privilege, Chain, ChainTy, Range, Dict, RD, Dom, Set, P, Univ, Schema, MovieSchema, RT,
        TDom, TRange, MoviesTable, RTSubEqFlds, RTEqFlds, Disjoint, Union, Ret, Seq, Trans, Map,
        Pure, Then, Else, Exit, Inv, OneState, Priv, Path, FH, Stable,
      Monoid, VerifiedMonoid, VerifiedComMonoid, Plus_2_2_eq_4, Plus_2_2, Nat_up, Int_up,
      AppendNilId, AppendAssoc,MapFusion,
      Plus_comm, Par, Term,HasTyPr,HasTyEv,IsSubTyPr,IsSubTyEv,StepPr,StepEv, StepsPr,StepsEv,Env,Prim,Expr,Proposition,DataProp,VName,ProofOf,ProofOfN,Type,Names,Kind,
      HasTy,WfType,IsSubTy,Step,Steps,EvalsTo,
      Formula, Assignment, Body, Accel, Real, Accel', RVar, VerifiedCommutativeMonoid, CommutativeMonoid
  },
  emphstyle={[2]\color{blue_violet}},
  emph=
  {[3]
    case,class,newtype,data,deriving,do,else,if,unpack,import,in,infixl,infixr,instance,let,
    module,of,primitive,then,refinement,type,where,forall,bound,
    measure,reflect,predicate, instance, class,
    exists
  },
  emphstyle={[3]\color{castanho_ulisses}\textbf},
  emph=
  {[4]
    quot,rem,div,mod,elem,notElem,seq
  },
  emphstyle={[4]\color{preto_ulisses}},
  emph=
  {[5]
    EQ,GT,LT,Left,Right,SBase,TPrim,TSub,TAbs,TAbsEx,TAbsCQ,
    TRefn,TInt,EApp,EPrim,EEq,EVar,PInt,Empty,Cons,EVar,ELam,EApp,TFunc,
    Refl,AddStep
  },
  emphstyle={[5]\color{haskell_green}},
  emph=
  {[6]
      axiomatize, measure, inline
  },
  emphstyle={[6]\color{lcolor}\ttfamily\itshape},
  emph=
  {[6]
      hasTyEvPr,isSubTyEvPr
  }
}

\lstnewenvironment{code}
{\lstset{language=HaskellUlisses}}
{}

\lstnewenvironment{mcode}
{\lstset{language=HaskellUlisses,columns=fullflexible,keepspaces,mathescape}}
{}

\lstnewenvironment{mcodef}
{\lstset{language=HaskellUlisses,columns=fullflexible,keepspaces,mathescape,frame=single}}
{}

\lstMakeShortInline[language=HaskellUlisses,mathescape,keepspaces,mathescape,basicstyle=\ttfamily\codesize,breakatwhitespace]@

\lstdefinelanguage{Pseudo} {
  basicstyle=\ttfamily\codesize,
  sensitive=true,
  mathescape=true,
  morecomment=[l][\color{gray_ulisses}\ttfamily\codesize]{--},
  morecomment=[s][\color{gray_ulisses}\ttfamily\codesize]{\{-}{-\}},
  morestring=[b]",
  showstringspaces=false,
  numberstyle=\codesize,
  numberblanklines=true,
  showspaces=false,
  breaklines=true,
  showtabs=false
}

\lstdefinelanguage{java} {
    keywordstyle=[1],
    keywordstyle=[2]\color{ForestGreen},
    keywordstyle=[3]\color{Bittersweet},
    keywordstyle=[4]\color{RoyalPurple},
    morekeywords={region,private,synchronized}
}

\ifdefined\withcolor
        \definecolor{typecol}{rgb}{0.0,0.5,0.0}
        \definecolor{funcol}{rgb}{0.0,0.1,0.9}
\else
        \definecolor{typecol}{gray}{0.0}
        \definecolor{funcol}{gray}{0.0}
\fi

\lstdefinelanguage{pseudo2}{
  language=Python,
  basicstyle=\ttfamily\normalsize,
  mathescape=true,
  morekeywords={type,def,do,let,unpack},
  emph={[1] \Expr,\Pred, HP, FP, Int},
  emphstyle={[1]\itshape\color{typecol}},
  emph={[2] \pbe,normalize},
  emphstyle={[2]\itshape\color{funcol}},
  emph={[3] repeat,until},
  emphstyle={[3]\textbf}
}

\lstnewenvironment{code2}{\lstset{language=pseudo2}}{}

\lstnewenvironment{fscode}
{\lstset{language=HaskellUlisses,basicstyle=\ttfamily\small,mathescape=true}}
{}

\lstnewenvironment{codebox}
{\lstset{language=HaskellUlisses,frame=tlbr,basicstyle=\ttfamily\codesize,mathescape=true}}
{}

\def \ha {\lstinline[language=HaskellUlisses,basicstyle=\ttfamily\incodesize,mathescape=true]}

\usepackage{semantic}
\usepackage{float}
\usepackage{commands_full}
\usepackage[shortcuts]{extdash}
\usepackage[capitalize]{cleveref}
\usepackage{enumitem}

\newtheorem{requirement}{Requirement}

\newlength{\hatchspread}
\newlength{\hatchthickness}
\newlength{\hatchshift}
\newcommand{\hatchcolor}{}
\tikzset{hatchspread/.code={\setlength{\hatchspread}{#1}},
         hatchthickness/.code={\setlength{\hatchthickness}{#1}},
         hatchshift/.code={\setlength{\hatchshift}{#1}},
         hatchcolor/.code={\renewcommand{\hatchcolor}{#1}}}
\tikzset{hatchspread=3pt,
         hatchthickness=0.4pt,
         hatchshift=0pt,
         hatchcolor=black}
\pgfdeclarepatternformonly[\hatchspread,\hatchthickness,\hatchshift,\hatchcolor]
   {custom north west lines}
   {\pgfqpoint{\dimexpr-2\hatchthickness}{\dimexpr-2\hatchthickness}}
   {\pgfqpoint{\dimexpr\hatchspread+2\hatchthickness}{\dimexpr\hatchspread+2\hatchthickness}}
   {\pgfqpoint{\dimexpr\hatchspread}{\dimexpr\hatchspread}}
   {
    \pgfsetlinewidth{\hatchthickness}
    \pgfpathmoveto{\pgfqpoint{0pt}{\dimexpr\hatchspread+\hatchshift}}
    \pgfpathlineto{\pgfqpoint{\dimexpr\hatchspread+0.15pt+\hatchshift}{-0.15pt}}
    \ifdim \hatchshift > 0pt
      \pgfpathmoveto{\pgfqpoint{0pt}{\hatchshift}}
      \pgfpathlineto{\pgfqpoint{\dimexpr0.15pt+\hatchshift}{-0.15pt}}
    \fi
    \pgfsetstrokecolor{\hatchcolor}
    \pgfusepath{stroke}
   }

\pgfdeclarepatternformonly[\hatchspread,\hatchthickness,\hatchcolor]
  {custom horizontal lines} 
  {\pgfpointorigin}{\pgfqpoint{100pt}{1pt}}{\pgfqpoint{100pt}{\dimexpr\hatchspread}}%
  {%
    \pgfsetlinewidth{\hatchthickness}%
    \pgfpathmoveto{\pgfqpoint{0pt}{0.5pt}}%
    \pgfpathlineto{\pgfqpoint{100pt}{0.5pt}}%
    \pgfsetstrokecolor{\hatchcolor}
    \pgfusepath{stroke}%
  }%

\pgfdeclarepatternformonly[\hatchspread,\hatchthickness,\hatchcolor]
  {custom vertical lines}
  {\pgfpointorigin}{\pgfqpoint{1pt}{100pt}}{\pgfqpoint{\dimexpr\hatchspread}{100pt}}%
  {%
    \pgfsetlinewidth{\hatchthickness}%
    \pgfpathmoveto{\pgfqpoint{0.5pt}{0pt}}%
    \pgfpathlineto{\pgfqpoint{0.5pt}{100pt}}%
    \pgfsetstrokecolor{\hatchcolor}
    \pgfusepath{stroke}%
  }%

\pgfdeclarepatternformonly[\hatchspread,\hatchthickness,\hatchcolor]
  {custom crosshatch dots}
  {\pgfqpoint{-1pt}{-1pt}}{\pgfqpoint{2.5pt}{2.5pt}}{\pgfqpoint{\dimexpr\hatchspread+\hatchspread}{\dimexpr\hatchspread+\hatchspread}}%
  {%
    \pgfpathcircle{\pgfqpoint{0pt}{0pt}}{\hatchthickness}%
    \pgfsetstrokecolor{\hatchcolor}%
    \pgfusepath{stroke}%
    \pgfusepath{fill}%
  }%

\pgfdeclarepatternformonly[\hatchcolor]%
  {custom checkerboard}
  {\pgfpointorigin}{\pgfqpoint{4mm}{4mm}}{\pgfqpoint{4mm}{4mm}}%
  {%
    \pgfpathrectangle{\pgfpointorigin}{\pgfqpoint{2mm}{2mm}}%
    \pgfsetfillcolor{\hatchcolor}%
    \pgfusepath{fill}%
    \pgfpathrectangle{\pgfqpoint{2mm}{2mm}}{\pgfqpoint{2mm}{2mm}}%
    \pgfsetfillcolor{\hatchcolor}%
    \pgfusepath{fill}%
  }%

\makeatother

\title{Mechanizing Refinement Types}

\author{Michael Borkowski}
\affiliation{\institution{UC San Diego} \country{USA}}
\email{mborkows@eng.ucsd.edu}

\author{Niki Vazou}
\affiliation{\institution{IMDEA Software Institute} \country{Spain}}
\email{niki.vazou@imdea.org}

\author{Ranjit Jhala}
\affiliation{\institution{UC San Diego} \country{USA}}
\email{rjhala@eng.ucsd.edu}

\begin{abstract}
Practical checkers based on refinement types
use the combination of implicit semantic subtyping
and parametric polymorphism to simplify the specification
and automate the verification of sophisticated properties
of programs.
However, a formal meta-theoretic accounting
of the \emph{soundness} of refinement type
systems using this combination has proved elusive.
We present \sysrf, a core refinement calculus
that combines semantic subtyping and parametric
polymorphism.
We develop a metatheory for this calculus
and prove soundness of the type system.
Finally, we give a full mechanization
\emph{of} our metatheory using the refinement-type
based \lh as a proof checker, showing how
refinements can be used \emph{for} mechanization.
\end{abstract}

\begin{document}
\pgfdeclarelayer{background}
\pgfdeclarelayer{middleground}
\pgfdeclarelayer{foreground}
\pgfsetlayers{background,middleground,main,foreground}

\maketitle

\section{Introduction}\label{sec:intro}

Refinements constrain types with
logical predicates to specify new
concepts.
For example, the refinement type
$\tpos \defeq \breft{\tint}{\vv}{0 < v}$
describes \emph{positive} integers
and $\tnat \doteq \breft{\tint}{\vv}{0 \leq v}$
specifies natural numbers.
Refinements on types have been successfully
used to define sophisticated concepts
(\eg secrecy \cite{FournetCCS11},
 resource constraints \cite{Knoth20},
 security policies~\cite{STORM})
that can then be verified in programs
developed in various programming
languages like Haskell~\cite{Vazou14},
Scala~\cite{kuncak-stainless},
Racket~\cite{RefinedRacket} and
Ruby~\cite{rruby}.

%
The success of refinement types
relies on the combination of two
essential features.
First, \textit{implicit} semantic
subtyping uses semantic (SMT-solver based)
reasoning to automatically convert
the types of expressions without
troubling the programmer for explicit type
casts.
For example, consider a positive expression
$e : \tpos$ and a function expecting natural
numbers $f:\funcftype{\tnat}{\tint}$.
To type check the application $f\ e$,
the refinement type system will implicitly
convert the type of $e$ from \tpos to \tnat,
because $0 < v \Rightarrow 0 \leq v$
semantically holds.
Importantly, refinement types propagate
semantic subtyping inside type constructors
to, for example, treat function arguments in
a contravariant manner.
Second, \textit{parametric polymorphism}
allows the propagation of the refined types
through polymorphic function interfaces,
without the need for extra reasoning.
As a trivial example, once we have
established that $e$ is positive,
parametric polymorphism should let
us conclude that $f\ e : \tpos$ if,
for example, $f$ is the identity function
$f:\funcftype{a}{a}$.

As is often the case with useful ideas,
the engineering of practical tools has
galloped far ahead of the development
of the meta-theoretical foundations
for refinements with subtyping and
polymorphism.
In fact, such a development is difficult.
As \citet{SekiyamaIG17} observe, a na\"ive combination of
type variables and subtyping leads to unsoundness
because potentially contradicting refinements
can be lost at type instantiation.
Their suggested solution replaces semantic with
syntactic subtyping, which is significantly less
expressive.
Other recent formalizations of refinement types
either drop semantic subtyping~\cite{kuncak-stainless}
or polymorphism~\cite{flanagan06, newfstar}.

In this paper we present \sysrf, a core calculus
with a refinement type system that combines semantic
subtyping with refined polymorphic type variables.
We develop and establish the properties of \sysrf
with four concrete contributions.

\mypara{1. Reconciliation}
Our first contribution is a language
that combines refinements and polymorphism
in a way that ensures the metatheory remains sound
without sacrificing the expressiveness needed
for practical verification.
To this end, \sysrf introduces a kind
system that distinguishes the type
variables that can be soundly refined
(without the risk of losing refinements
at instantiation) from the rest,
which are then left unrefined.
In addition our design includes
a form of existential typing \cite{Knowles09}
which is essential to \emph{synthesize} the types
-- in the sense of bidirectional typing -- for applications
and let-binders in a compositional manner (\S~\ref{sec:language}, \ref{sec:lang:static}).

\mypara{2. Foundation}
Our second contribution is to establish the foundations
of \sysrf by proving soundness, which
says that if $e$ has a type then,
either $e$ is a value or it can
step to another term of the same type.
%
%
The combination of semantic subtyping, polymorphism,
and existentials makes the soundness proof challenging
with circular dependencies that do not arise in standard
(unrefined) calculi.
To ease the presentation and tease out the essential
ingredients of the proof we stage the metatheory.
First, we review an unrefined \emph{base} language
\sysf, a classic System F \cite{TAPL} with primitive
\tint and \tbool types (\S~\ref{sec:soundnessF}).
Next, we show how refinements (kinds, subtyping,
and existentials) must be accounted for to
establish the soundness of \sysrf (\S~\ref{sec:soundness}).

\mypara{3. Reification}
Our third contribution is to 
introduce \textit{data propositions} a novel feature
that enables the encoding of derivation trees for inductively
defined judgments as refined data types, by first reifying
the propositions and evidence as plain Haskell data, and then
using refinements to connect the two.
Hence, data propositions let us write plain Haskell functions
over refined data to provide explicit, constructive proofs (\S~\ref{sec:data-props}).
Without data propositions reasoning about 
potentially non-terminating computations was not possible in \lh,
thereby precluding even simple
meta-theoretic developments such as the soundness
of \sysf let alone \sysrf.

\mypara{4. Verification}
Our final contribution is to fully mechanize the metatheory
of \sysrf using \lh. 
Our formalization uses data propositions and recursive
Haskell functions on derivation
trees to produce explicit witnesses that correspond
to proofs of our soundness theorems~\cite{Vazou18}.
Our proof is non-trivial, requiring 9,400 lines
of code and 30 minutes to verify.
While we are certain that it is possible,
and perhaps simpler, to implement the mechanization with
specialized proof assistants and tactics,
our contribution is to show this is feasible
purely as a (refined) Haskell program.
Indeed, we show that substantial meta-theoretical
formalizations over arbitrary computations
are feasible via data propositions via
\lh-style refinement typing (\S~\ref{sec:implementation}).
\section{Overview}
\label{sec:overview}

Our overall strategy is to present the metatheory
for \sysrf in two parts.
First, we review the metatheory for \sysf:
a familiar starting point that corresponds to
the full language with refinements erased (\S~\ref{sec:soundnessF}).
Second, we use the scaffolding established
by \sysf to highlight the extensions needed to
develop the metatheory for refinements in
\sysrf (\S~\ref{sec:soundness}).
Let's begin with a high-level overview that describes
a proof skeleton that is shared across the
developments for \sysf and \sysrf,
the specific challenges posed by refinements,
and the machinery needed to go from the simpler
\sysf to the refined \sysrf.

\mypara{Types and Terms}
Both \sysf and \sysrf have the same syntax for \emph{terms} $e$
(\cref{fig:syn:terms}).
\sysf has the usual syntax for \emph{types} $t$ familiar from System F,
while \sysrf additionally allows (\sysf's) types to be \emph{refined}
by terms (respectively, the white parts and all of \cref{fig:syn:types}),
and existential types.
Both languages include a notion of \emph{kinds} $\kindex$ that
qualify the types that are allowed to be refined.

\mypara{Judgments}
Both languages have \emph{typing} judgments
$\hastype{\tcenv}{e}{t}$ which say that a
term $e$ has type $t$ with respect to a
binding environment (\ie context) $\tcenv$.
Additionally, both languages have \emph{well-formedness}
judgments $\isWellFormed{\tcenv}{t}{\kindex}$ which say
that a type $t$ has the kind $k$ in context $\tcenv$, by
requiring that the free variables in $t$ are appropriately
bound in the environment $\tcenv$.
(Though some presentations of \sysf \cite{TAPL}
eschew well-formedness judgments, they are helpful for
a mechanized metatheory \cite{AydemirCPPW08}).
Crucially, \sysrf has a \emph{subtyping} judgment
$\isSubType{\tcenv}{t_1}{t_2}$ which says that type
$t_1$ is a subtype of $t_2$ in context $\tcenv$.
Subtyping for refined base types is established
via an axiomatized 
\emph{implication} judgment
$\imply{\tcenv}{p}{q}$ which says that the term
$p$ logically implies the term $q$ whenever their
free variables are given values described by $\tcenv$.
We take an axiomatized approach to capture the
properties required by an implication checking oracle
for proving soundness.

\mypara{Proof Landscape}
\cref{fig:graph} charts the overall
landscape of our formal development as a
dependency graph of the main lemmas which
establish meta-theoretic properties of the
different judgments.
Nodes shaded \colboth represent lemmas
in the metatheories for \sysf and \sysrf.
The \colref nodes denote lemmas
that only appear in 
\sysrf.
An arrow shows a dependency: the lemma at the
\emph{tail} is used in the proof of the lemma at
the \emph{head}. 
Darker arrows are dependencies in \sysrf only.

\definecolor{palegrey}{rgb}{0.72, 0.72, 0.72}
\definecolor{lightgreen}{rgb}{0.92, 0.92, 0.92}
\definecolor{medgreen}{rgb}{0.02, 0.02, 0.02}
\definecolor{darkgreen}{rgb}{0.02, 0.02, 0.02}
\definecolor{sysfcolor}{rgb}{0.52, 0.52, 0.52}
\begin{figure}[t]
\begin{center}
\scalebox{0.64}{
    \begin{tikzpicture}[
        > = stealth, 
        auto,
        node distance = 1.0cm,         
        semithick                      
    ]

    \tikzstyle{every state}=[
        rectangle,
        minimum width=3.70cm,          
        draw=black, rounded corners
    ]

    %

    \node[state, fill=palegrey]   (we1)                    {Weaken: tv in sub};
    \node[state, fill=lightgreen] (we2) [above of=we1]     {Weaken: tv in typ};
    \node[state, fill=palegrey]   (we3) [above of=we2]     {Weaken: var in sub};
    \node[state, fill=lightgreen] (we4) [above of=we3]     {Weaken: var in typ};

    \node[state, fill=lightgreen] (we5) [right=5.40cm of we1] {Weaken: tv in wf};
    \node[state, fill=lightgreen] (we6) [right=5.40cm of we2] {Weaken: var in wf};

    \node[state, fill=palegrey]   (su1) [above right=-0.25cm and 0.75cm of we2] {Substitute: tv in sub};
    \node[state, fill=lightgreen] (su2) [above of=su1]     {Substitute: tv in typ};
    \node[state, fill=palegrey]   (su3) [above of=su2]     {Substitute: var in sub};
    \node[state, fill=lightgreen] (su4) [above of=su3]     {Substitute: var in typ};

    \node[state, fill=lightgreen] (su5) [above=1.00cm of we6] {Substitute: tv in wf};
    \node[state, fill=lightgreen] (su6) [above of=su5] {Substitute: var in wf};

    \begin{pgfonlayer}{background}
        \path (su4.west |- su4.north)+(-0.25,0.60) node (oa) {};
        \path (su1.south -| su1.east)+(0.25,-0.25) node (oc) {};
        \path[rounded corners, draw=black!50, dashed]
            (oa) rectangle (oc) node (orangefill) {};
        \fill[pattern=custom north west lines,hatchspread=6pt,hatchthickness=1pt,hatchcolor=gray!40]
            (oa) rectangle (oc) ;
    \end{pgfonlayer}
    \begin{pgfonlayer}{background}
        \path (we4.west |- we4.north)+(-0.25,0.60) node (oa1) {};
        \path (we1.south -| we1.east)+(0.25,-0.25) node (oc1) {};
        \path[rounded corners, draw=black!50, dashed]
            (oa1) rectangle (oc1) node (orangefill) {};
        \fill[pattern=custom north west lines,hatchspread=6pt,hatchthickness=1pt,hatchcolor=gray!40]
            (oa1) rectangle (oc1) ;
    \end{pgfonlayer}

    \begin{pgfonlayer}{middleground}
        \path (we4.west |- we4.north)+(-0.10,0.45) node (b3a) {};
        \path (we1.south -| we1.east)+(0.10,-0.10) node (b3b) {};
        \path (we4.north) +(0,0.20) node (b3cap) {Weakening Lemma};
        \path[draw=black!70]
            (b3a) rectangle (b3b) node (b3) {};
    \end{pgfonlayer}
    \begin{pgfonlayer}{middleground}
        \path (we6.west |- we6.north)+(-0.10,0.45) node (b4a) {};
        \path (we5.south -| we5.east)+(0.10,-0.10) node (b4b) {};
        \path (we6.north) +(0,0.20) node (b4cap) {Weakening Lemma};
        \path[draw=black!70]
            (b4a) rectangle (b4b) node (b4) {};
    \end{pgfonlayer}
    \begin{pgfonlayer}{middleground}
        \path (su4.west |- su4.north)+(-0.10,0.45) node (b5a) {};
        \path (su1.south -| su1.east)+(0.10,-0.10) node (b5b) {};
        \path (su4.north) +(0,0.20) node (b5cap) {Substitution Lemma};
        \path[draw=black!70]
            (b5a) rectangle (b5b) node (b5) {};
    \end{pgfonlayer}
    \begin{pgfonlayer}{middleground}
        \path (su6.west |- su6.north)+(-0.10,0.45) node (b6a) {};
        \path (su5.south -| su5.east)+(0.10,-0.10) node (b6b) {};
        \path (su6.north) +(0,0.20) node (b6cap) {Substitution Lemma};
        \path[draw=black!70]
            (b6a) rectangle (b6b) node (b6) {};
    \end{pgfonlayer}

    \begin{pgfonlayer}{foreground}
        \path[every edge/.style={draw, ->, >={Stealth[width = 5pt, length = 7pt, inset=1pt,sep]}}]
            (we5.north east) edge[sysfcolor, bend right] node {} (su6.south east)
            (we2.north east) edge[sysfcolor, bend left] node {} (su3.south west)
            (we5.north west) edge[sysfcolor, bend left] node {} (we3.south east)
            (su5.north west) edge[sysfcolor, bend right] node {} (su3.south east);
    \end{pgfonlayer}
    \begin{pgfonlayer}{foreground}
        \path[every edge/.style={draw, >-<, >={Stealth[width = 6pt, length = -5pt, inset=-8pt]}}]
            (we1) edge[darkgreen, bend right=15] node {} (we2)
            (we3) edge[darkgreen, bend right=15] node {} (we4)
            (su1) edge[darkgreen, bend right=15] node {} (su2)
            (su3) edge[darkgreen, bend right=15] node {} (su4);
    \end{pgfonlayer}

    \node[state, fill=palegrey]   (na1) [left=5.25cm of su1]   {Exact Subtypes};
    \node[state, fill=palegrey]   (na2) [above=0.75cm of na1]   {Exact Types};
    \node[state, fill=palegrey]   (na3) [above=0.75cm of na2]   {Narrowing Lemmas};

    \begin{pgfonlayer}{background}
        \path (na3.west |- na3.north)+(-0.25,0.25) node (ra) {};
        \path (na1.south -| na1.east)+(0.25,-0.25) node (rc) {};
        \path[rounded corners, draw=black!50, dashed]
            (ra) rectangle (rc) node (redfill) {};
        \fill[pattern=custom checkerboard, hatchcolor=gray!25]
            (ra) rectangle (rc);
    \end{pgfonlayer}

    \begin{pgfonlayer}{foreground}
        \path[every edge/.style={draw, ->, >={Stealth[width = 5pt, length = 7pt, inset=1pt,sep]}}]
            (na1.north) edge[medgreen] node {} (na2.south)
            (na2.north) edge[medgreen] node {} (na3.south)
            (na2.east) edge[medgreen, bend left=45] node {} (su2.north west);
    \end{pgfonlayer}

    \node[state, fill=palegrey]   (in1) [above left=-0.25cm and 0.80cm of su4]   {Transitivity};
    \node[state, fill=lightgreen] (in2) [above=0.60cm of in1]   {Inversion of Typing};

    \begin{pgfonlayer}{background}
        \path (in2.west |- in2.north)+(-0.25,0.25) node (ya) {};
        \path (in1.south -| in1.east)+(0.25,-0.25) node (yc) {};
        \path[rounded corners, draw=black!50, dashed]
            (ya) rectangle (yc) node (yellowfill) {};
        \fill[pattern=custom vertical lines, hatchspread=5pt, hatchthickness=1pt, hatchcolor=gray!35]
            (ya) rectangle (yc);
    \end{pgfonlayer}
    \begin{pgfonlayer}{foreground}
        \path[every edge/.style={draw, ->, >={Stealth[width = 5pt, length = 7pt, inset=1pt,sep]}}]
            (in1.north) edge[medgreen] node {} (in2.south)
            (na3.east) edge[medgreen] node {} (in2.west)
            (na3.east) edge[medgreen] node {} (in1.west)
            (su4.west) edge[medgreen] node {} (in1.east);
    \end{pgfonlayer}

    \node[state, fill=lightgreen] (pr1) [left=0.75cm of in2]   {Canonical Forms};
    \node[state, fill=lightgreen] (pr2) [right=0.75cm of in2]   {Primitives};
    \node[state, fill=lightgreen] (pr3) [right=0.75cm of pr2]   {Polym. Prim.};
    \node[state, fill=lightgreen] (pr4) [above left=1.15cm and -0.50cm of in2]   {Progress};
    \node[state, fill=lightgreen] (pr5) [right=1.80cm of pr4]   {Preservation};
    \node[state, fill=lightgreen] (pr6) [above right=-0.25 cm and 2.20cm of pr5]   {Values Stuck};
    \node[state, fill=lightgreen] (pr7) [below=0.17cm of pr6]   {Det. Semantics};

    \begin{pgfonlayer}{foreground}
        \path[every edge/.style={draw, ->, >={Stealth[width = 5pt, length = 7pt, inset=1pt,sep]}}]
            (in2.north) edge[sysfcolor] node {} (pr4.south)
            (pr1.north) edge[sysfcolor, bend left] node {} (pr4.south west)
            (su4.north) edge[medgreen] node {} (pr2.south)
            (su4.east) edge[medgreen] node {} (pr3.south)
            (pr1.north east) edge[sysfcolor] node {} (pr5.south west)
            (in2.north) edge[sysfcolor] node {} (pr5.south)
            (pr2.north) edge[sysfcolor] node {} (pr4.south)
            (pr2.north) edge[sysfcolor] node {} (pr5.south)
            (su4.north east) edge[sysfcolor, bend right=55] node {} (pr5.south east)
            (pr3.north) edge[sysfcolor] node {} (pr4.south east)
            (pr3.north) edge[sysfcolor] node {} (pr5.south east)
            (pr4.east) edge[sysfcolor] node {} (pr5.west)
            (pr6.west) edge[sysfcolor] node {} (pr5.east)
            (pr7.west) edge[sysfcolor] node {} (pr5.east);
    \end{pgfonlayer}

    \end{tikzpicture}
}
\end{center}
\caption{Logical dependencies in the metatheory. We write
``var'' to abbreviate a term variable and ``tv'' to abbreviate a type variable.}
\label{fig:graph}
\end{figure}
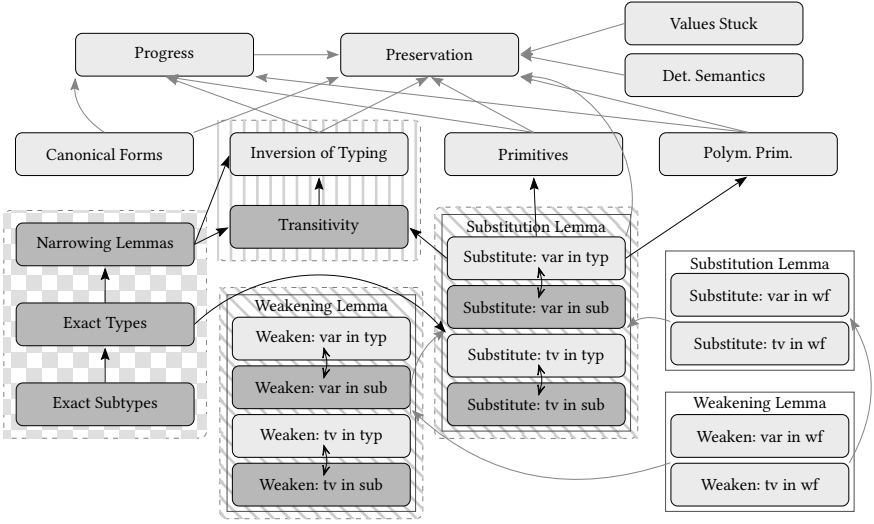

%

\mypara{Soundness via Preservation and Progress}
For both \sysrf and \sysf we establish soundness via
\begin{itemize}[leftmargin=*]
\item \emphbf{Progress:}
      If a closed term is well-typed,
      then either it is a value
      or it can be further evaluated;
\item \emphbf{Preservation:}
      If a closed term is well-typed,
      then its type is preserved
      under evaluation.
\end{itemize}
The type soundness theorem states  that a well-typed closed term
cannot become {\em stuck}: any sequence of evaluation steps will
either end with a value or the sequence can be extended
by another step.
Next, we describe the lemmas used
to establish preservation and progress
for \sysf and then outline the essential
new ingredients that demonstrate soundness
for the refined \sysrf.


\subsection{Metatheory for \sysf}

\mypara{Progress} in \sysf is standard
as the typing rules are syntax-directed.
The top-level rule used to obtain the typing derivation
for a term $e$ uniquely determines the syntactic structure
of $e$ which lets us use the appropriate small-step reduction
rule to obtain the next step of the evaluation of $e$.

\mypara{Preservation} says that when a well-typed
expression $e$ steps to $e'$, then $e'$ is also well-typed.
As usual, the non-trivial case is when the step is
a type abstraction $\tabs{\alpha}{\kindex}{e}$
(respectively lambda abstraction $\vabs{x}{e}$)
\emph{applied to}  a type (respectively value),
in which case the term $e'$ is obtained by substituting
the type or value appropriately in $e$.
Thus, our \sysf metatheory requires us to prove
a \emph{Substitution Lemma},
which describes how typing judgments behave under
substitution of free term or type variables.
Additionally, some of our typing rules use
well-formedness judgments and so we must also
prove that well-formedness is preserved by substitution.

\mypara{Substitution} requires some technical lemmas
that let us weaken judgments by adding any fresh variable
to the binding environment.

\mypara{Primitives}
Finally, the primitive reduction steps
(\eg arithmetic operations) require the assumption
that the reduction rules defined for the
built-in primitives are type preserving.
%

\subsection{What's hard about Refinements?}
\label{overview:exists}

\mypara{Subtyping}
Refinement types rely on implicit semantic subtyping,
that is, type conversion (from subtypes) happens without
any explicit casts and is checked semantically via logical
validity.
For example, consider a function $f$ that requires natural
numbers as input, applied to a positive argument $e$.
Let
$$
    \tcenv \doteq f:\funcftype{\tnat}{\tint}, e:\tpos
$$
The application $f\ e$ will type check as below,
using the \tSub rule to implicitly convert the
type of the argument and the \sBase rule to
check that positive integers are always naturals
by checking the validity of the formula $\forall v.\ 0 < v \Rightarrow 0 \leq v$.
$$
\inferrule*[Right=\tApp]{
    \inferrule*[Right=\tVar]{ }{
    \hastype{\tcenv}{f}{\funcftype{\tnat}{\tint}}
    }
    \qquad
    \qquad
    \qquad
    \inferrule*[Right=\tSub]{
        \inferrule*[Right=\tVar]{ }{
            \hastype{\tcenv}{e}{\tpos}
        }
        \qquad\qquad
        \inferrule*[Right=\sBase]{ \forall v.\ 0 < v \Rightarrow 0 \leq v }{
              \isSubType{\tcenv}{\tpos}{\tnat}
        }
     }{
        \hastype{\tcenv}{e}{\tnat}
    }
 }{
    \hastype{\tcenv}{\app{f}{e}}{\tint}
}
$$

Importantly, most refinement type systems
use type-constructor directed rules to destruct
subtyping obligations into basic (semantic)
implications.
For example, in~\cref{fig:subtyping}
the rule \sFunc states that functions
are covariant on the result and contravariant
on the arguments.
Thus, a refinement type system can,
without any annotations or casts,
decide that $e : \funcftype{\tnat}{\tpos}$
is a suitable argument for the higher order function
$f : \funcftype{(\funcftype{\tpos}{\tnat})}{\tint}$.

\mypara{Existentials}
For compositional and decidable type checking, some
refinement type systems use an existential type~\cite{flanagan-exist} to check
dependent function application, \ie the \rulename{TApp-Exists} rule below, instead
of the standard type-theoretic \rulename{TApp-Exact} rule.
$$
\inferrule*[Right=\rulename{TApp-Exact}]{
    \hastype{\tcenv}{f}{\functype{x}{\stype_x}{t}} \qquad \hastype{\tcenv}{e}{\stype_x}
 }{
    \hastype{\tcenv}{\app{f}{e}}{\stype\subst{}{x}{e}}
} \qquad\qquad\qquad
\inferrule*[Right=\rulename{TApp-Exists}]{
    \hastype{\tcenv}{f}{\functype{x}{\stype_x}{t}} \qquad \hastype{\tcenv}{e}{\stype_x}
 }{
    \hastype{\tcenv}{\app{f}{e}}{\existype{x}{\stype_x}{\stype}}
}
$$

To understand the difference, consider
some expression $e$ of type $\tpos$ and the identity function $f$
\begin{align*}
    e&:\tpos & f&:\functype{x}{\tint}{\breft{\tint}{v}{v = x}}
\end{align*}
The application $f\ e$ is typed as
$\breft{\tint}{v}{v = e}$ with the \rulename{TApp-Exact} rule,
which has two problems.
First, the information that $e$ is positive is lost.
To regain this information the system needs to re-analyze the expression $e$
breaking compositional reasoning. %
Second, the arbitrary expression $e$ enters the refinement logic
making it impossible for the system to restrict refinements into decidable
logical fragments.
Using the \rulename{TApp-Exists} rule, both of these problems are addressed.
The type of $f\ e$ becomes
$\existype{x}{\tpos}{\breft{\tint}{v}{v = x}}$
preserving the information that the application argument is positive,
while the variable $x$ cannot break any carefully crafted decidability guarantees.

\citet{flanagan-exist} introduce the existential application rule
and show that it preserves the decidability and completeness of the refinement type system.
An alternative approach for decidable and compositional type checking
is to ensure that all the application arguments
are variables by ANF transforming the original
program~\cite{Flanagan93}.
ANF is more amicable to \emph{implementation}
as it does not require the definition of one
more type form.
However, ANF is more problematic for the
\emph{metatheory}, as ANF is not preserved
by evaluation.
Additionally, existentials let us \emph{synthesize}
types for let-binders in a bidirectional style: when
typing $\eletin{x}{e_1}{e_2}$, the existential lets
us eliminate $x$ from the type synthesized for
$e_2$, yielding a precise, algorithmic
system \cite{CosmanICFP17}.
Thus, we choose to use existential types in \sysrf.

\mypara{Polymorphism}
Polymorphism is a precious type
abstraction~\cite{10.1145/99370.99404},
but combined with refinements, it can lead to
imprecise or, worse, unsound systems.
As an example, below we present the
function \ha{max} with four potential
type signatures.
$$\begin{array}{rrrcl}
& \text{Definition} &  \texttt{max} &  = & \lambda x\ y . \texttt{if } x < y \texttt{ then } y \texttt{ else } x \\[0.05in]
\text{Attempt 1:}& \textit{Monomorphism} & \texttt{max} & :: &
\functype{x}{\tint}{\functype{y}{\tint}{\breft{\tint}{\vv}{x \leq \vv \wedge y \leq \vv}}}\\
\text{Attempt 2:}& \textit{Unrefined Polymorphism} & \texttt{max} &  :: &
\functype{x}{\al}{\functype{y}{\al}{\al}}\\
\text{Attempt 3:}& \textit{Refined Polymorphism} & \texttt{max} &  :: &
\functype{x}{\al}{\functype{y}{\al}{\breft{\al}{\vv}{x \leq \vv \wedge y \leq \vv}}}\\
\sysrf\text{:} & \textit{Kinded Polymorphism} & \texttt{max} &  :: &
\polytype{\al}{\skbase}\functype{x}{\al}{\functype{y}{\al}{\breft{\al}{\vv}{x \leq \vv \wedge y \leq \vv}}}
\end{array}$$
As a first attempt, we give \ha{max} a monomorphic type,
stating that the result of  \ha{max} is an integer greater
or equal to any of its arguments.
This type is insufficient because it forgets any information known for
\ha{max}'s arguments. For example, if both arguments are positive,
the system cannot decide that \ha{max x y} is also positive.
To preserve the argument information we give \ha{max} a polymorphic
type, as a second attempt. Now the system can deduce that
\ha{max x y} is positive, but forgets that it is also
greater or equal to both \ha{x} and \ha{y}.
In a third attempt, we naively combine the benefits of
polymorphism with refinements to give \ha{max} a very precise type
that is sufficient to propagate the arguments' properties (positivity)
and \ha{max} behavior (inequality).

Unfortunately, refinements on arbitrary type
variables are dangerous for two reasons.
First, the type of \ha{max} implies
that the system allows comparison
between any values (including functions).
Second, if refinements on type variables
are allowed, then, for soundness~\cite{Belo11},
all the types that substitute variables should be refined.
For example, if a type variable
is refined with \tfalse (that is, $\breft{\al}{\vv}{\tfalse}$)
and gets instantiated with an unrefined function
type ($\functype{x}{t_x}{t}$),
then the \tfalse refinement
is lost and the system becomes unsound.

\mypara{Base Kind when Refined}
To preserve the benefits on refinements
on type variables, without the complications
of refining function types, we introduce
a kind system that separates the type
variables that can be refined with the
ones that cannot.
Variables with the base kind $\skbase$,
can be refined, compared, and only
substituted by base, refined types.
The other type variables have kind $\skstar$
and can only be trivially refined with \ttrue.
With this kind system, we give \ha{max}
a polymorphic and precise type that
naturally rejects non comparable
(\eg function) arguments.

\subsection{From \sysf to \sysrf}\label{sec:overview:rf}

The metatheory for \sysrf requires us to enrich
that of \sysf with three essential and non-trivial
blocks --- shown as shaded regions in \cref{fig:graph} ---
that help surmount the challenges posed by the combination
of refinements with existentials, subtyping and polymorphism.

\mypara{Typing Inversion}
First, thanks to (refinement) subtyping \sysrf is not
syntax directed, and so we cannot directly invert the
typing derivations of terms to get derivations for their
sub-terms.
For example, we cannot directly invert a derivation
$\hastype{\tcenv}{\vabs{x}{e}}{\functype{x}{t_x}{t}}$
to obtain a typing derivation that the body $e$ has
type $t$ because the above derivation may have been
established using (multiple instances of) subtyping.
The typing inversion lemmas addresses this problem by
using the \emph{transitivity of subtyping} to restructure
the judgment tree to collapse all use of subtyping
in a way that lets us invert the non-subtyping judgment
to conclude that if a term (\eg $\vabs{x}{e}$)
is well-typed, then its components (\eg $e$)
are also well-typed.
The proof of transitivity of subtyping is non-trivial
due to the presence of existential types. We cannot
proceed by induction on the structure of the two
subtyping judgments 
($\isSubType{\tcenv}{\stype_1}{\stype_2}$
and $\isSubType{\tcenv}{\stype_2}{\stype_3}$), 
because we do not apply the inductive hypothesis
directly to subderivations. We must first apply the
substitution and narrowing lemmas in various cases,
which may increase the size of derivations used
in the inductive hypothesis (\S~\ref{sec:soundness:inversion}). 
Instead,  our proof goes by induction
on a more intricate size (the 
combined depth of types $\stype_1$, $\stype_2$, and $\stype_3$).

\mypara{Subtyping}
The biggest difference between the two
metatheories is that \sysrf has a notion
of subtyping which is crucial to making
refinements practical.
Subtyping complicates \sysrf by introducing
a mutual dependency between the lemmas for
typing and subtyping judgments.
Recall that typing depends on subtyping
due to the usual subsumption rule (\tSub in~\cref{fig:t}) that lets
us weaken the type of a term with a super-type.
Conversely, subtyping depends upon typing because
of the rule (\textsc{S-Witn} in~\cref{fig:s})
which establishes subtyping between \emph{existential}
types.
Thanks to this mutual dependency,
all of the lemmas from \sysf that relate
to typing judgments, \ie the weakening
and substitution lemmas, 
are now mutually recursive with new versions
for subtyping judgments shown in the \colsubtyping
region in~\cref{fig:graph}.

\mypara{Narrowing}
Finally, due to subtyping, the proofs of the
typing inversion and substitution lemmas for \sysrf
require \emph{narrowing} lemmas that
allow us to replace a type that appears
inside the binding environment of a judgment
with a subtype, thus ``narrowing'' the scope
of the judgment.
Due to the mutual dependencies
between the typing and subtyping judgments
of \sysrf, we must prove narrowing
for both typing and subtyping, which in turn depend
on narrowing for well-formedness judgments.
%
%
A few important cases of the narrowing
proofs require other technical lemmas
shown in the \colnarrowing region of \cref{fig:graph}.
For example, the proof of narrowing for
the ``occurrence-typing'' rule \textsc{T-Var} that
crucially enables path-sensitive reasoning,
uses a lemma on \emph{selfifying} \cite{Ou2004}
the types involved in the judgments.

\section{Language}
\label{sec:language}

\begin{figure}[t!]
\begin{tabular}{rrcll}
\emphbf{Primitives} 
  & \sconst & $\bnfdef$ & $\ttrue$ $\spmid$ $\tfalse$           & \emph{booleans} \\
  &         & $\spmid$  & $0, 1, 2, \ldots$                     & \emph{integers} \\ 
  &         & $\spmid$  & $\wedge, \vee, \neg, \leftrightarrow$ & \emph{boolean ops.} \\
  &         & $\spmid$  & $\leq, =$                             & \emph{polymorphic comparisons} \\ [0.05in] 

\emphbf{Values}
  & \sval   & $\bnfdef$ & \sconst               & \emph{primitives} \\ 
  &         & $\spmid$  & x, y, \ldots          & \emph{variables} \\
  &         & $\spmid$  & \vabs{x}{e}           & \emph{abstractions} \\
  &         & $\spmid$  & \tabs{\alpha}{k}{e}   & \emph{type abstractions} \\[0.05in]

\emphbf{Terms}
  & \sexpr  & $\bnfdef$ & \sval                 & \emph{values} \\ 
  &         & $\spmid$  & \app{e_1}{e_2}        & \emph{applications} \\
  &         & $\spmid$  & \tyapp{e}{t}          & \emph{type applications} \\
  &         & $\spmid$  & \eletin{x}{e_1}{e_2}  & \emph{let-binders} \\
  &         & $\spmid$  & \tyann{e}{t}          & \emph{annotations} \\
\end{tabular}
\caption{Syntax of Primitives, Values, and Expressions.}
\label{fig:syn:terms}
\end{figure}

For brevity, clarity and also 
to cut a circularity in the metatheory
(in rule \wtRefn in \S~\ref{sec:typing:wf}), 
we formalize 
refinements using two calculi. 
%
%
The first is the \emph{base} language 
\sysf: a classic System F \cite{TAPL} 
with call-by-value semantics extended 
with primitive \tint and \tbool types 
and operations.
The second calculus is the \emph{refined} language
\sysrf which extends \sysf with refinements.
By using the first calculus to express the typing judgments
for our refinements,
we avoid making the well-formedness and typing judgments be
mutually dependent in our full language.
We use the $\greybox{\mbox{grey}}$ highlights 
to indicate the extensions to the syntax and 
rules of $\sysf$ needed to support refinements 
in $\sysrf$.

\begin{figure}[t!]
  \begin{tabular}{rrcll}
  \emphbf{Kinds} 
    & \skind & $\bnfdef$ & \skbase & \emph{base kind} \\
    &        & $\spmid$  & \skstar & \emph{star kind} \\[0.05in]
  
  \emphbf{Predicates}  
    & \spred & $\bnfdef$ & $\greybox{\{ e \;|\; \exists\, \tcenv.\, \hasftype{\tcenv}{\sexpr}{\tbool} \}}$ & \greytextbox{\emph{boolean-typed terms}} \\[0.05in] 
  
  \emphbf{Base Types} 
    & \sbase & $\bnfdef$ & $\tbool$ & \emph{booleans} \\
    &        & $\spmid$  & $\tint$  & \emph{integers} \\
    &        & $\spmid$  & $\tvar$  & \emph{type variables} \\[0.05in]
  
  \emphbf{Types}
    & \stype & $\bnfdef$ & $\sbase\greybox{\!\breft{}{\vv}{\spred}}$  & \emph{\greytextbox{refined} base type} \\ 
    &        & $\spmid$  & $\greybox{x\!}\!\functype{}{t_x}{t}$         & \emph{function type}     \\        
    &        & $\spmid$  & \greybox{\existype{x}{t_x}{t}} & \greytextbox{\emph{existential type}}  \\        
    &        & $\spmid$  & $\polytype{\tvar}{\skind}{t}$  & \emph{polymorphic type}  \\ [0.05in]        
  
  \emphbf{Environments}
    & $\tcenv$ & $\bnfdef$ & $\varnothing$                  & \emph{empty environment} \\
    &          & $\spmid$  & $\tcenv, \bind{x}{t}$          & \emph{variable binding} \\
    &          & $\spmid$  & $\tcenv, \bind{\tvar}{\skind}$ & \emph{type binding} \\
  \end{tabular}
  \caption{Syntax of Types. 
           The grey boxes are the extensions 
           to $\sysf$ needed by $\sysrf$.
           We use $\sftype$ for $\sysf$-only types.}
  \label{fig:syn:types}
  \label{fig:syn:reft}
  \label{fig:syn:env}
  \end{figure}

\subsection{Syntax} \label{sec:lang:syntax}

We start by describing the syntax of terms and types 
in the two calculi.  

\mypara{Constants, Values and Terms}
\cref{fig:syn:terms} summarizes the 
syntax of terms in both calculi.
Terms are stratified into primitive \emph{constants} 
and \emph{values}.
The primitives $\sconst$
include \tint and \tbool constants, 
primitive boolean operations, 
and polymorphic comparison and equality 
primitive.
%
%
Values $\sval$ are those expressions 
which cannot be evaluated any further, 
including primitive constants, binders 
and $\lambda$- and type- abstractions.
Finally, the terms $\sexpr$ comprise values,
value- and type- applications, let-binders 
and annotated expressions.

\mypara{Kinds \& Types}       
\cref{fig:syn:types} shows the syntax of the types,
with the grey boxes indicating the extensions to $\sysf$ 
required by $\sysrf$.
In \sysrf, only base types \tbool and \tint 
can be refined: we do not permit refinements 
for functions and polymorphic types. 
\sysrf enforces this restriction using two kinds
which denote types that may ($\skbase$) or may not ($\skstar$)
be refined.
The (unrefined) \emph{base} types $\sbase$ comprise 
$\tint$, $\tbool$, and type variables $\tvar$. 
The simplest type is of the form
$\breft{\sbase}{\vv}{\spred}$ 
comprising a base type $\sbase$ 
and a \emph{refinement} that restricts 
$\sbase$ to the subset of values 
$\vv$ that satisfy $\spred$ \ie 
for which $\spred$ evaluates to $\ttrue$.
We use refined base types to build up
dependent function types (where the input 
parameter $x$ can appear in the output type's 
refinement), existential and polymorphic 
types.
%
In the sequel, we write $\sbase$ to abbreviate 
$\breft{\sbase}{\vv}{\ttrue}$
and call types refined with only \ttrue 
``trivially refined'' types. 

\mypara{Refinement Erasure}
The reduction semantics of our polymorphic 
primitives are defined 
using an \emph{erasure} function that 
returns the unrefined, \sysf version of a refined \sysrf type:
\[
  \forgetreft{ \breft{\sbase}{\vv}{\spred} } \defeq \sbase, \quad 
  \forgetreft{ \functype{x}{t_x}{t} } \defeq \forgetreft{t_x} \rightarrow \forgetreft{t}, \quad
  \forgetreft{ \existype{x}{t_x}{t} } \defeq \forgetreft{t}, \quad{\rm and} \quad
  \forgetreft{ \polytype{\al}{k}{t} } \defeq \polytype{\al}{k}{\forgetreft{t}}
\]

\mypara{Environments}
\cref{fig:syn:env} describes 
the syntax of typing environments 
$\tcenv$ which contain both term 
variables bound to types and type 
variables bound to kinds. 
These variables may appear in types 
bound later in the environment.
In our formalism, environments grow 
from right to left.

\mypara{Note on Variable Representation}
Our metatheory requires that all variables 
bound in the environment be distinct. 
Our mechanization enforces this invariant 
via the locally nameless representation~\cite{Aydemir05}: 
free and bound variables become distinct objects 
in the syntax, as are type and term variables.
All free variables have unique names which
never conflict with bound variables represented as 
de Bruijn indices. This eliminates
the possibility of capture in substitution
and the need to perform alpha-renaming during
substitution. 
The locally nameless representation avoids 
technical manipulations such as index shifting by using 
names instead of indices for  free variables 
(we discuss alternatives in~\S~\ref{sec:related}).
To simplify the presentation of the syntax and rules, we 
use names for bound variables to make the dependent nature
of the function arrow clear.

\subsection{Dynamic Semantics} \label{sec:lang:dynamic}
\begin{figure}
  \judgementHead{Operational Semantics}{$\sexpr \step \sexpr'$}
        \begin{mathpar} 
        \inferrule*[Right=\ePrim]{  }{\app{\sconst}{\sval} \step \delta(\sconst,\sval)} \and
        \inferrule*[Right=\eTPrim]{  }{\tyapp{\sconst}{\stype} \step \delta_T(\sconst,\forgetreft{\stype})} \and
        \inferrule*[Right=\eAnn]{\sexpr \step \sexpr'}{\tyann{\sexpr}{\stype} \step \tyann{\sexpr'}{\stype}}\and
        \inferrule*[Right=\eAnnV]{ }{\tyann{\sval}{\stype} \step \sval}\\
        \inferrule*[Right=\eApp]{\sexpr \step \sexpr'}{\app{\sexpr}{\sexpr_1} \step \app{\sexpr'}{\sexpr_1}} \and
        \inferrule*[Right=\eAppV]{\sexpr \step \sexpr'}{\app{\sval}{\sexpr} \step \app{\sval}{\sexpr'}} \and
        \inferrule*[Right=\eAppAbs]{ }
          {\app{(\vabs{x}{\sexpr})}{\sval} \step \subst{\sexpr}{x}{\sval}} \\
        \inferrule*[Right=\eTApp]{\sexpr \step \sexpr'}{\tyapp{\sexpr}{t} \step \tyapp{\sexpr'}{t}} \and
        \inferrule*[Right=\eTAppAbs]{ }
          {\tyapp{(\tabs{\tvar}{\skind}{\sexpr})}{\stype} \step \subst{\sexpr}{\tvar}{\stype}} \\ 
        \inferrule*[Right=\eLet]{ \sexpr_x \step \sexpr'_x}
          {\eletin{x}{\sexpr_x}{\sexpr} \step \eletin{x}{\sexpr'_x}{\sexpr}} \and
        \inferrule*[Right=\eLetV]{ }{\eletin{x}{\sval}{\sexpr} \step \subst{\sexpr}{x}{\sval}} 
        \end{mathpar}
  \caption{The small-step semantics.} 
  \label{fig:e}
  \label{fig:opsem}
\end{figure}

\cref{fig:opsem} summarizes the substitution-based, 
call-by-value, contextual, small-step semantics 
for both calculi.
We specify the reduction semantics 
of the primitives using the functions 
$\delta$ and $\delta_T$.

\begin{figure}[t!]
$$\begin{array}{r@{\hskip 0.05in}c@{\hskip 0.05in}l}
\subst{\breft{\tvarb}{x}{p}}{\al}{t_\al}       & \defeq & \breft{\tvarb}{x}{\subst{p}{\al}{t_\al}}, \al \not = \tvarb \\ 
\subst{(\functype{x}{t_x}{t})}{\al}{t_\al}     & \defeq & \functype{x}{(\subst{t_x}{\al}{t_\al})}{\subst{t}{\al}{t_\al}} \\ 
\subst{(\existype{x}{t_x}{t})}{\al}{t_\al}     & \defeq & \existype{x}{(\subst{t_x}{\al}{t_\al})}{\subst{t}{\al}{t_\al}} \\ 
\subst{(\polytype{\tvarb}{k}{t})}{\al}{t_\al}  & \defeq & \polytype{\tvarb}{k}{\subst{t}{\al}{t_\al}} \\
\subst{\breft{\al}{x}{p}}{\al}{t_\al}       & \defeq & \strengthen{t_\al}{\subst{p}{\al}{t_\al}}{x} \\
& & \\
\strengthen{\breft{\al}{z}{q}}{p}{x}              & \defeq & \breft{\al}{z}{\subst{p}{x}{z} \wedge q} \\
\strengthen{\existype{z}{t_z}{t}}{p}{x}           & \defeq & \existype{z}{t_z}{\strengthen{t}{p}{x}} \\
\strengthen{\functype{x}{t_x}{t}}{\_}{\_}           & \defeq & \functype{x}{t_x}{t} \\
\strengthen{\polytype{\al}{k}{t}}{\_}{\_}           & \defeq & \polytype{\al}{k}{t} \\
\end{array}$$
\caption{Type substitution.}
\label{fig:type-subst}
\end{figure}

\mypara{Substitution} 
%
The key difference with standard formulations
is the notion of substitution for type variables 
at (polymorphic) type-application sites as shown 
in rule $\eTAppAbs$ in~\cref{fig:opsem}.
\cref{fig:type-subst} summarizes how type 
substitution is defined, which is standard except 
for the last line which defines the substitution 
of a type variable $\al$ in a refined type variable 
${\breft{\al}{x}{p}}$ with a type \stype which is potentially 
refined.
To do this substitution, we combine $p$ with the type \stype 
by using $\strengthen{\stype}{p}{x}$ 
which essentially conjoins the refinement $p$ 
to the top-level refinement of a base-kinded 
$\stype$. 
%
For existential types, $\mathsf{strengthen}$ 
\emph{pushes} the refinement through the 
existential quantifier. 
Function and quantified types are left unchanged 
as they cannot be used to instantiate a \emph{refined} 
type variable (which must be of base kind).

\mypara{Primitives}
The function $\delta(\sconst, \sval)$ 
specifies what an application $\app{\sconst}{\sval}$ 
of a built-in monomorphic primitive evaluates to.
The reductions are defined in a curried 
manner, \ie we have that $\app{\app{\leq}{m}}{n} \steps \delta(\delta(\leq,m),n)$. 
Currying gives us unary relations like $m\!\!\leq$ 
which is a partially evaluated version of the $\leq$ relation.
We also denote by $\delta_T(=, \forgetreft{\stype})$
and $\delta_T(\leq, \forgetreft{\stype})$ 
a function specifying the reduction rules for type 
applications for the polymorphic 
built-in primitives $=$ and $\leq$.
$$\begin{array}{rclrclrcl}
\delta(\wedge,{\tt true}) & \defeq & \lambda x.\, x &
  \delta(\leftrightarrow,{\tt true}) & \defeq & \lambda x.\, x & 
    \delta_T(=, \tbool) & \defeq & \leftrightarrow  \\
\delta(\wedge,{\tt false}) & \defeq & \lambda x.\, {\tt false} &
  \delta(\leftrightarrow,{\tt false}) & \defeq & \lambda x.\, \neg x &
    \delta_T(=, \tint) & \defeq & = \\
\delta(\vee,{\tt true}) & \defeq & \lambda x.\, {\tt true} &
  \delta(\leq,m) & \defeq & m\!\!\leq  & 
  \delta_T(\leq, \tbool) & \defeq & \leq  \\
\delta(\vee,{\tt false}) & \defeq & \lambda x.\, x &
  \delta(m\!\!\leq, n) & \defeq & {\tt}(m \leq n) &
  \delta_T(\leq, \tint) & \defeq & \leq  \\
\delta(\neg,{\tt true}) & \defeq & {\tt false} & 
  \delta(=,m) & \defeq & m\!\!= &
    &  & \\
\delta(\neg,{\tt false}) & \defeq &  {\tt true} &
  \delta(m\!\!=, n) & \defeq &  {\tt}(m = n) &
    & & \\
\end{array}$$
         


\mypara{Determinism}
Our proof of soundness uses the following 
property of the operational semantics.
\begin{lemma}[Determinism]\label{lem:step-determ}
For every expression $\sexpr$, 
\begin{itemize} 
    \item there exists at most one term $\sexpr'$ such that $\sexpr \step \sexpr'$, 
    \item there exists at most one value $\sval$ such that $\evalsTo{\sexpr}{\sval}$, and
    \item if $\sexpr$ is a value there is no term $\sexpr'$ such that $\sexpr \step \sexpr'$.
\end{itemize}
\end{lemma}    

\section{Static Semantics}
\label{sec:lang:static}

The static semantics of our calculi comprise
four main judgment forms:
{\emph{well-formedness}} judgments that determine when a type
or environment is syntactically well-formed (in $\sysf$ and $\sysrf$);
{\emph{typing}} judgments that stipulate that a term has
a particular type in a given context (in $\sysf$ and $\sysrf$);
{\emph{subtyping}} judgments that establish when one type can
be viewed as a subtype of another (in $\sysrf$); and
{\emph{implication}} judgments that establish when one predicate
implies another (in $\sysrf$).
Next, we present the static semantics of \sysrf by describing
each of these judgments and the rules used to establish them.
We use $\greybox{\mbox{grey}}$ to highlight the antecedents and rules
specific to $\sysrf$.

\mypara{Co-finite Quantification}
We define our rules using the co-finite quantification technique
of~\citet{AydemirCPPW08}.
This technique enforces a small (but critical) restriction
in the way fresh names are introduced in the antecedents of rules.
For example, below we present the standard (on the left)
and our (on the right) rules for type abstraction.
$$
\inferrule*[Right=\tAbsEx]
    {\notmem{\al'}{\tcenv} \;\;
      \hastype{\bind{\al'}{\skind},\tcenv}{\subst{\sexpr}{\al}{\al'}}{\subst{\stype}{\al}{\al'}} }
    {\hastype{\tcenv}{\tabs{\al}{\skind}{\sexpr}}{\polytype{\al}{\skind}{\stype}}}
\qquad\qquad
\inferrule*[Right=\tTAbs]
    {\forall\notmem{\al'}{L}. \;\;
      \hastype{\bind{\al'}{\skind},\tcenv}{\subst{\sexpr}{\al}{\al'}}{\subst{\stype}{\al}{\al'}} }
    {\hastype{\tcenv}{\tabs{\al}{\skind}{\sexpr}}{\polytype{\al}{\skind}{\stype}}}
$$
%
The standard rule \tAbsEx requires the \text{existence} of a
fresh type variable name $\al'$.
Instead our co-finite quantification rule
states that the rule holds for any name excluding
a finite set of names $L$. 
As observed by~\citet{AydemirCPPW08} this rephrasing
simplifies the mechanization of metatheory
by eliminating the need for renaming lemmas.

\subsection{Well-formedness}
\label{sec:typing:wf}

\mypara{Judgments}
The ternary judgment $\isWellFormed{\tcenv}{\stype}{\skind}$
says that the type $\stype$ is well-formed in the environment
$\tcenv$ and has kind $\skind$.
The judgment $\isWellFormedE{\tcenv}$ says that the
environment $\tcenv$ is well formed, meaning
that variables are only bound to well-formed
types.
Well-formedness is also used in the (unrefined) system $\sysf$, where
$\isWFFT{\tcenv}{\sftype}{\skind}$
means that the (unrefined) $\sysf$ type
$\sftype$ is well-formed in environment
$\tcenv$ and has kind $\skind$
and $\isWFFE{\tcenv}$ means
that the free type and expression variables
of the unrefined
environment $\tcenv$ are bound earlier in the environment.
While well-formedness is not strictly
required for \sysf, we found it helpful
to simplify the mechanization \cite{Remy21}.

\mypara{Rules}
\cref{fig:wf} summarizes the rules
that establish the well-formedness of types
and environments, with the grey highlighting
the parts relevant for refinements.
Rule \wtBase states that the two closed base
types (\tint and \tbool) are well-formed and have
base kind on their own or with trivial refinement \ttrue.
Similarly, rule \wtVar says that an unrefined or trivially refined
type variable $\tvar$ is well-formed having kind $\skind$
so long as $\bind{\tvar}{\skind}$ is bound in the environment.
The rule \wtRefn stipulates that a refined base type $\breft{\sbase}{x}{\spred}$
is well-formed with base kind in some environment
if the unrefined base type $\sbase$
has base kind in the same environment and if
the refinement predicate $\spred$ has type $\tbool$
in the environment augmented by binding a fresh variable to type $\sbase$.
Note that if $\sbase \equiv \tvar$ then we can only form the antecedent
$\isWellFormed{\tcenv}{\breft{\tvar}{x}{\ttrue}}{\skbase}$
when $\bind{\tvar}{\skbase} \in \tcenv$ (rule \wtVar),
which prevents us from refining star-kinded type variables.
%
To break a circularity in our judgments,
in which well-formedness judgments can
appear in the antecedent position of
typing judgments and a typing judgment
would appear in the antecedent position
of \wtRefn, we stipulate only a $\sysf$
judgment for $\spred$ having underlying
type $\tbool$.
Our rule \wtFunc states that a function
type $\functype{x}{t_x}{t}$ is well-formed
with star kind in some environment $\tcenv$
if both type $t_x$ is well-formed (with any kind)
in the same environment and type $t$
is well-formed (with any kind) in the
environment $\tcenv$ augmented by binding
a fresh variable to $t_x$.
Rule \wtExis states that an existential
type $\existype{x}{t_x}{t}$ is well-formed
with some kind $\skind$ in some environment
$\tcenv$ if both type $t_x$ is well-formed
(with any kind) in the same environment
and type $t$ is well-formed with kind
$\skind$ in the environment $\tcenv$
augmented by binding a fresh variable
to $t_x$.
Rule \wtPoly establishes that a
polymorphic type $\polytype{\tvar}{\skind}{\stype}$
has star kind in environment $\tcenv$ if the
inner type $\stype$ is well-formed (with any kind)
in environment $\tcenv$ augmented by binding a fresh
type variable $\tvar$ to kind $\skind$.
Finally, rule \wtKind simply states that if a type $\stype$
is well-formed with base kind in some environment, then
it is also well-formed with star kind. 
This rule is required by our metatheory to convert base to star
kinds in type variables.

As for environments, rule \wfeEmp states that the empty environment
is well-formed. Rule \wfeBind says that a well-formed environment
$\tcenv$ remains well-formed after binding a fresh variable $x$ to any
type $t_x$ that is well-formed in $\tcenv$.
Finally rule \wfeTBind states that a well-formed environment remains
well-formed after binding a fresh type variable to any kind.

\begin{figure}[t!]
\judgementHead{Well-formed Type}{\isWellFormed{\tcenv}{\stype}{\skind}}

\begin{mathpar}
%
\inferrule*[Right=\wtBase]
    {\sbase \in \{\tbool, \tint\}}
    {\isWellFormed{\tcenv}{\sbase\greybox{\!\breft{}{x}{\ttrue}}\!}{\skbase}}
\and
\inferrule*[Right=\wtVar]
    {\bind{\al}{\skind} \in \tcenv}
    {\isWellFormed{\tcenv}{\al\greybox{\!\breft{}{x}{\ttrue}}\!}{\skind}} \\
\greybox{\inferrule*[Right=\;\;\wtRefn]
    { \isWellFormed{\tcenv}{\breft{b}{x}{\ttrue}}{\skbase} \quad
      \forall\notmem{y}{\tcenv}. \quad
      \hasftype{\bind{y}{b}, \forgetreft{\tcenv}}{\subst{p}{x}{y}}{\tbool}
    }
    {\isWellFormed{\tcenv}{\breft{b}{x}{p}}{\skbase}} } \\
\inferrule*[Right=\wtFunc]
    { \isWellFormed{\tcenv}{t_x}{\skind_x} \quad
      \greybox{\forall\notmem{y}{\tcenv}.} \quad
      \isWellFormed{\greybox{\bind{y}{t_x},}\tcenv}{t\greybox{\!\subst{}{x}{y}}}{\skind}
    }
    { \isWellFormed{\tcenv}{\greybox{x\!}\!\functype{}{t_x}{t}}{\skstar} } \\
\greybox{\inferrule*[Right=\;\;\wtExis]
    { \isWellFormed{\tcenv}{t_x}{\skind_x} \quad
      \forall\notmem{y}{\tcenv}. \quad
      \isWellFormed{\bind{y}{t_x},\tcenv}{\subst{t}{x}{y}}{\skind}
    }
    { \isWellFormed{\tcenv}{\existype{x}{t_x}{t}}{\skind} } }\\
\inferrule*[Right=\wtPoly]
    { \forall\notmem{\al'}{\tcenv}. \quad
      \isWellFormed{\bind{\al'}{k}, \tcenv}{\subst{t}{\al}{\al'}}{\skind_t}
    }
    { \isWellFormed{\tcenv}{\polytype{\al}{\skind}{t}}{\skstar} }
\and
\inferrule*[Right=\wtKind]
    { \isWellFormed{\tcenv}{t}{\skbase} }
    { \isWellFormed{\tcenv}{t}{\skstar} } \\
\end{mathpar}

\judgementHead{Well-formed Environment}{\isWellFormedE{\tcenv}}

\begin{mathpar}
\inferrule*[Right=\wfeEmp]
    { }
    { \isWellFormedE{\varnothing} }
\and
\inferrule*[Right=\wfeBind]
    {\isWellFormed{\tcenv}{\stype_x}{\skind_x} \quad
     \isWellFormedE{\tcenv} \quad
     \notmem{x}{\tcenv}
    }
    { \isWellFormedE{\bind{x}{\stype_x}, \tcenv} }
\and
\inferrule*[Right=\wfeTBind]
    {\isWellFormedE{\tcenv} \quad
     \notmem{\al}{\tcenv}
    }
    { \isWellFormedE{\bind{\al}{\skind}, \tcenv} }
\end{mathpar}

\caption{Well-formedness of types and environments. The rules for
  $\sysf$ exclude the grey boxes.}
\label{fig:wf}
\label{fig:wfe}
\end{figure}

\subsection{Typing}
\label{sec:typing:typ}

The judgment $\hastype{\tcenv}{\sexpr}{\stype}$ states
that the term $\sexpr$ has type $\stype$ in the context of
environment $\tcenv$.
We write $\hasftype{\tcenv}{\sexpr}{\sftype}$
to indicate that term $\sexpr$ has the (unrefined)
$\sysf$ type $\sftype$ in the (unrefined) context
$\tcenv$.
\cref{fig:typing} summarizes
the rules that establish typing for both $\sysf$ and
$\sysrf$, with grey 
for the $\sysrf$ extensions.

\begin{figure}
  \judgementHead{Typing}{\hastype{\tcenv}{\sexpr}{\stype}}
    \begin{mathpar}             
        \inferrule*[Right=\tPrim]
        {\ty{\sconst} = \stype}{\hastype{\tcenv}{\sconst}{\stype}}
    \and
        \inferrule*[Right=\tVar]
        {\bind{x}{\stype} \in \tcenv  \qquad
         \greybox{\isWellFormed{\tcenv}{\stype}{\skind}}}
        {\hastype{\tcenv}{x}{\greybox{\self{\whitebox{\stype}}{x}{\skind}}}}
    \and
        \inferrule*[Right=\tApp]
        {\hastype{\tcenv}{\sexpr}{\greybox{x\!}\!\functype{}{\stype_x}{\stype}} \qquad
         \hastype{\tcenv}{\sexpr_x}{\stype_x}}
        {\hastype{\tcenv}{\app{\sexpr}{\sexpr_x}}{\greybox{\existype{x}{\stype_x}{\whitebox{\stype}}}}} \\
        \inferrule*[Right=\tAbs]
        { \forall\notmem{y}{\tcenv}.\quad
          \hastype{\bind{y}{\stype_x},\tcenv}{\subst{\sexpr}{x}{y}}{\greybox{\subst{\whitebox{\stype}}{x}{y}}} \qquad
         \isWellFormed{\tcenv}{\stype_x}{\skind_x}
         }
        {\hastype{\tcenv}{\vabs{x}{\sexpr}}{\greybox{x\!}\!\functype{}{\stype_x}{\stype}}}
      \and
        \inferrule*[Right=\tTApp]
        {\hastype{\tcenv}{\sexpr}{\polytype{\al}{\skind}{s}} \qquad
         \isWellFormed{\tcenv}{t}{\skind}}
        {\hastype{\tcenv}{\tyapp{\sexpr}{t}}{\subst{s}{\al}{t}}} \\
        \inferrule*[Right=\tTAbs]
        {\forall\notmem{\al'}{\tcenv}.\quad
        \hastype{\bind{\al'}{\skind},\tcenv}{\subst{\sexpr}{\al}{\al'}}{\subst{\stype}{\al}{\al'}}
        }
        {\hastype{\tcenv}{\tabs{\al}{\skind}{\sexpr}}{\polytype{\al}{\skind}{\stype}}} \\
        \inferrule*[Right=\tLet]
        {\hastype{\tcenv}{\sexpr_x}{\stype_x} \qquad
        \forall\notmem{y}{\tcenv}.\quad
        \hastype{\bind{y}{\stype_x},\tcenv}{\subst{\sexpr}{x}{y}}
             {\stype\greybox{\!\subst{}{x}{y}}}  \qquad
         \greybox{\isWellFormed{\tcenv}{\stype}{\skind}}
         }
        {\hastype{\tcenv}{\eletin{x}{\sexpr_x}{\sexpr}}{\stype}} \\
        \inferrule*[Right=\tAnn]
        {\hastype{\tcenv}{\sexpr}{\stype} \qquad
         \greybox{\isWellFormed{\tcenv}{\stype}{\skind}}}
        {\hastype{\tcenv}{\tyann{\sexpr}{\stype}}{\stype}}
    \and
        \greybox{
          \inferrule*[Right=\;\;\tSub]
          {\hastype{\tcenv}{\sexpr}{s} \qquad
           \isSubType{\tcenv}{s}{t} \qquad
           \isWellFormed{\tcenv}{t}{\skind}}
          {\hastype{\tcenv}{\sexpr}{t}}
        }
    \end{mathpar}
\caption{Typing rules.
\label{fig:typing}
The judgment $\hasftype{\tcenv}{\sexpr}{\sftype}$ is defined by excluding the grey boxes.}\label{fig:t}\label{fig:typing}
\end{figure}

\mypara{Typing Primitives}
The type of a built-in primitive $\sconst$ is given by
the function $\ty{\sconst}$, which is defined for every
constant of our system. Below we present essential
examples of the $\ty{\sconst}$ definition.
$$\begin{array}{rcl}
\ty{\ttrue} & \defeq & \breft{\tbool}{x}{x = \ttrue} \\
\ty{3} & \defeq& \breft{\tint}{x}{x = 3} \\
\ty{\wedge} & \defeq & \functype{x}{\tbool}{\functype{y}{\tbool}{\breft{\tbool}{v}{v = x \wedge y}}} \\
\ty{m\!\!\leq} & \defeq & \functype{y}{\tint}{\breft{\tbool}{v}{v = (m \leq y)}} \\
\ty{\leq} & \defeq & \polytype{\al}{\skbase}{\functype{x}{\al}{\functype{y}{\al}{\breft{\tbool}{v}{v = (x \leq y)}}}} \\
\ty{=} & \defeq & \polytype{\tvar}{\skbase}{\functype{x}{\al}{\functype{y}{\al}{\breft{\tbool}{v}{v = (x = y)}}}}
\end{array}$$
We note that the $=$ used in the refinements is the polymorphic
equals with type applications elided.
Further, we use $m\!\!\leq$ to represent
an arbitrary member of the infinite family
of primitives $0\!\!\leq,\, 1\!\!\leq,\, 2\!\!\leq,\ldots$.
For \sysf we erase the refinements
using $\forgetreft{\ty{\sconst}}$.
The rest of the definition is similar.

Our choice to make the typing and reduction
of constants external to our language,
\ie respectively given by the functions
$\ty{\sconst}$ and $\tc{\sconst}$,
makes our system easily extensible with further constants.
The requirement, for soundness, is that
these two functions on constants
together satisfy the following four conditions.
%
\begin{requirement}(Primitives) \label{lem:prim-typing}
For every primitive $c$,
\begin{enumerate}
\item If $\ty{\sconst} = \breft{\sbase}{x}{\spred}$, then
  $\isWellFormed{\varnothing}{\ty{\sconst}}{\skbase}$ and
  $\imply{\varnothing}{\ttrue}{\subst{\spred}{x}{\sconst}}$.
\item If $\ty{\sconst} = \functype{x}{\stype_x}{\stype}$ or
         $\ty{\sconst} = \polytype{\al}{\skind}{\stype}$, then
         $\isWellFormed{\varnothing}{\ty{\sconst}}{\skstar}$.
\item If ${\ty{\sconst}} = \functype{x}{\stype_x}{\stype}$,
      then for all $v_x$ such that
     $\hastype{\varnothing}{v_x}{\stype_x}$,
    $\hastype{\varnothing}{\delta(\sconst,v_x)}{\subst{\stype}{x}{v_x}}$.
\item If ${\ty{\sconst}} = \polytype{\al}{\skind}{\stype}$,
      then for all $\stype_\al$ such that
     $\isWellFormed{\varnothing}{\stype_\al}{\skind}$,
    $\hastype{\varnothing}{\delta_T(\sconst,\stype_\al)}{\subst{\stype}{\al}{\stype_\al}}$.
\end{enumerate}
\end{requirement}

To type constants, rule \tPrim gives the type
$ty(\sconst)$ to any built-in
primitive $\sconst$, in any context.
The typing rules for boolean and integer
constants are included in \tPrim.

\mypara{Typing Variables with Selfification}
Rule \tVar establishes that any variable $x$ that
appears as $\bind{x}{\stype}$ in environment $\tcenv$
can be given the \emph{selfified} type \cite{Ou2004}
$\self{\stype}{x}{\skind}$ provided that
$\isWellFormed{\tcenv}{\stype}{\skind}$.
This rule is crucial in practice,
to enable path-sensitive ``occurrence'' typing \cite{Tob08},
where the types of variables are refined by control-flow guards.
For example, suppose we want to establish
$\hastype{\bind{\al}{\skbase}}{(\vabs{x}{x})}{\functype{x}{\al}{\breft{\al}{y}{x=y}}}$,
and not just $\hastype{\bind{\al}{\skbase}}{(\vabs{x}{x})}{\funcftype{\al}{\al}}$.
The latter judgment would result
from applying rule \tAbs
if \tVar merely stated that
$\hastype{\tcenv}{x}{\stype}$
whenever $\bind{x}{\stype} \in \tcenv$.
%
%
Thus we need to strengthen the \tVar rule to
be \emph{selfified}.
%
Informally, to get information about $x$
into the refinement level, we need to say
that $x$ is constrained to elements of
type $\al$ that are equal to $x$ itself.
In order to express the exact type of
variables, we introduce a ``selfification''
function that strengthens a refinement
with the condition that a value is equal
to itself.
Since abstractions do not admit equality,
we only selfify the base types and the existential
quantifications of them;
using the $\selfname$ function defined below.
$$\begin{array}{r@{\hskip 0.03in}c@{\hskip 0.03in}l@{\hskip 0.1in}r@{\hskip 0.03in}c@{\hskip 0.03in}l}
\self{\breft{\sbase}{z}{p}}{x}{\skbase} & \defeq & \breft{\sbase}{z}{p \wedge z = x} &
  \self{\existype{z}{t_z}{t}}{x}{\skind} & \defeq & \existype{z}{t_z}{\self{t}{x}{\skind}} \\
\self{\functype{x}{t_x}{t}}{\_}{\_} & \defeq & \functype{x}{t_x}{t} &
  \self{\polytype{\al}{k}{t}}{\_}{\_} & \defeq & \polytype{\al}{k}{t}
\end{array}$$

\mypara{Typing Applications with Existentials}
Our rule \tApp states the conditions for typing
a term application $\app{\sexpr}{\sexpr_x}$.
Under the same environment,
we must be able to type $\sexpr$ at
some function type $\functype{x}{\stype_x}{\stype}$ and
$\sexpr_x$ at $\stype_x$. Then we can give $\app{\sexpr}{\sexpr_x}$
the existential type $\existype{x}{\stype_x}{\stype}$.
%
The use of existential types in rule \tApp
is one of the distinctive features of our language
and was introduced by~\citet{Knowles09}.
As overviewed in \S~\ref{overview:exists},
we chose this form of \tApp over the conventional
form with $\hastype{\tcenv}{\app{\sexpr}{\sexpr_x}}{\subst{\stype}{x}{\sexpr_x}}$
in the consequent position because our version prevents
the substitution of arbitrary expressions (\eg functions and type abstractions)
into refinements.
As an alternative, we could have used A-Normal Form~\cite{Flanagan93},
but since this form is not preserved under the small step operational semantics, it
would greatly compicate our metatheory, by forcing the definition of
closing subtitutions for non-value expressions.

\mypara{Other Typing Rules}
Rule \tAbs says that we can type a lambda abstraction
$\vabs{x}{\sexpr}$ at a function type $\functype{x}{\stype_x}{\stype}$
whenever $\stype_x$ is well-formed and the body $\sexpr$
can be typed at $\stype$ in the environment augmented by binding
a fresh variable to $\stype_x$.
Our rule \tTApp states that whenever a term $\sexpr$ has polymorphic
type $\polytype{\al}{\skind}{s}$, then for any well-formed type $\stype$
with kind $\skind$ in the same environment, we can give the type
$\subst{s}{\al}{\stype}$ to the type application $\tyapp{\sexpr}{\stype}$.
For the \sysf variant of \tTApp, we erase the refinements (via $\forgetreft{\stype}$)
before checking well-formedness and performing the substitution.
The rule \tTAbs establishes that a type-abstraction $\tabs{\al}{\skind}{\sexpr}$
can be given a polymorphic type $\polytype{\al}{\skind}{\stype}$ in some
environment $\tcenv$ whenever $\sexpr$ can be given the well-formed type $\stype$
in the environment $\tcenv$ augmented by binding a fresh type variable
to kind $\skind$.
Next, rule $\tLet$ states that an expression $\eletin{x}{\sexpr_x}{e}$ has type
$\stype$ in some environment whenever $\stype$ is well-formed, $\sexpr_x$ can
be given some type $\stype_x$, and the body $\sexpr$ can be given type $\stype$
in the augmented environment formed by binding a fresh variable to $\stype_x$.
Rule $\tAnn$ establishes that an explicit annotation $\tyann{\sexpr}{\stype}$
can indeed be given the type $\stype$ when the underlying expression has
type $\stype$ and $\stype$ is well-formed in the same context.
The \sysf version of the rule erases the refinements and uses
$\forgetreft{\stype}$.
Finally, rule \tSub tells us that we can exchange a subtype $s$
for a supertype $t$ in a judgment $\hastype{\tcenv}{s}{t}$
provided that
$t$ is well-formed in the
same context and
$\isSubType{\tcenv}{s}{t}$, which we present next.
\subsection{Subtyping}
\label{sec:typing:sub}

\mypara{Judgments}
\cref{fig:s} summarizes the rules
that establish the subtyping judgment.
The \emph{subtyping} judgment ${\isSubType{\tcenv}{s}{t}}$
stipulates that the type $s$ is a subtype of
type the $t$ in the environment $\tcenv$
and is used in the subsumption typing rule \tSub (of~\cref{fig:typing}).
%

\mypara{Subtyping Rules}
%
%
The rule \sFunc states that one function type $\functype{x_1}{t_{x1}}{t_1}$
is a subtype of another function type $\functype{x_2}{t_{x2}}{t_2}$ in a
given environment $\tcenv$ when both $t_{x2}$ is a subtype of $t_{x1}$
and $t_1$ is a subtype of $t_2$ when we augment $\tcenv$ by
binding a fresh variable to type $t_{x2}$.
As usual, note that function subtyping is contravariant
in the input type and covariant in the outputs.
Rules \sBind and \sWitn establish subtyping for existential
types \cite{Knowles09}, respectively when the existential
appears on the left or right.
Rule \sBind allows us to exchange a universal quantifier
(a variable bound to some type $\stype_x$ in the environment)
for an existential quantifier.
If we have a judgment of the form
$\isSubType{\bind{y}{\stype_x},\tcenv}{\subst{\stype}{x}{y}}{\stype'}$
where $y$ does \emph{not} appear free in either $\stype'$ or in the context $\tcenv$,
then we can conclude that $\existype{x}{\stype_x}{\stype}$
is a subtype of $\stype'$.
Rule \sWitn states that if type $\stype$ is a subtype of
$\subst{\stype'}{x}{\sval_x}$ for some value $\sval_x$
of type $\stype_x$, then we can discard the specific
\emph{witness} for $x$ and quantify existentially to
obtain that $\stype$ is a subtype of $\existype{x}{\stype_x}{\stype'}$.
Rule \sPoly states when one polymorphic type
$\polytype{\al}{\skind}{\stype_1}$
is a subtype of another polymorphic type
$\polytype{\al}{\skind}{\stype_2}$
in some environment $\tcenv$. The requirement is that
$\stype_1$ be a subtype of $\stype_2$ in the
environment where we augment $\tcenv$ by binding
a fresh type variable to kind $\skind$.

Refinements enter the scene in the rule \sBase which
uses implication to specify
that a refined base type $\breft{\sbase}{x_1}{p_1}$
is a subtype of another $\breft{\sbase}{x_2}{p_2}$
in context $\tcenv$ when $p_1$ \emph{implies} $\spred_2$
in the environment $\tcenv$ augmented by binding
a fresh variable to the unrefined type $\sbase$.
Next, we describe how
implication is formalized in our system.

\subsection{Implication}
\label{sec:typing:implication}

The \emph{implication} judgment $\imply{\tcenv}{\spred_1}{\spred_2}$
states that the implication $\spred_1 \Rightarrow \spred_2$
is (logically) valid under the assumptions captured by the context $\tcenv$.
In refinement type implementations~\cite{newfstar,Seidel14}, this relation
is implemented as an external automated (usually SMT) solver.
In non-mechanized refinement type formalizations, there have been two approaches
to formalize predicate implication.
Either directly reduce it into a logical implication (\eg in~\citet{Gordon2010PrinciplesAA})
or define it using operational semantics (\eg in~\citet{Vazou18}).
It turns out that none of these approaches
can be directly encoded in a mechanized proof.
The former approach is insufficient because
it requires a formal connection between the (deeply embedded)
terms of \sysrf and the terms of the logic, which has not yet been clearly
established.
The second approach is more direct, since it gives meaning
to implication using directly the terms of \sysrf, via denotational semantics.
Sadly, the definition of denotational semantics for our
polymorphic calculus is not currently possible:
encoding type denotations as an inductive data type
(or proposition in our \lh encoding~\S~\ref{sec:implementation})
requires a negative occurrence which is not currently admitted.

\mypara{Abstracting over SMT-based Implication}
To bypass these problems we follow the approach
of~\citet{LehmannTanter} and encode implication
as an axiomatized judgment that satisfies the
below requirements.


\begin{requirement}\label{lem:implication}
  The implication relation satisfies the following statements:
  \begin{enumerate}
      \item (Reflexivity) $\imply{\tcenv}{\spred}{\spred}$.
      \item (Transitivity) If $\imply{\tcenv}{\spred_1}{\spred_2}$
        and $\imply{\tcenv}{\spred_2}{\spred_3}$, then
        $\imply{\tcenv}{\spred_1}{\spred_3}$.
      \item (Introduction) If $\imply{\tcenv}{\spred_1}{\spred_2}$
        and $\imply{\tcenv}{\spred_1}{\spred_3}$, then
        $\imply{\tcenv}{\spred_1}{\csand{\spred_2}{\spred_3}}$.
      \item (Conjunction 1)
      $\imply{\tcenv}{\csand{\spred_1}{\spred_2}}{\spred_1}$.
      \item (Conjunction 2)
      $\imply{\tcenv}{\csand{\spred_1}{\spred_2}}{\spred_2}$.
      \item (Repetition)
      $\imply{\tcenv}{\csand{\spred_1}{\spred_2}}
                     {\csand{\spred_1}{\csand{\spred_1}{\spred_2}}}$.
      \item (Narrowing) If
      $\imply{\tcenv_1,\bind{x}{t_x},\tcenv_2}{\spred_1}{\spred_2}$
      and $\isSubType{\tcenv_2}{s_x}{t_x}$, then
      $\imply{\tcenv_1,\bind{x}{s_x},\tcenv_2}{\spred_1}{\spred_2}$.
      \item (Weakening) If
      $\imply{\tcenv_1,\tcenv_2}{\spred_1}{\spred_2}$,
       $a,x\not\in\tcenv$, then
      $\imply{\tcenv_1,\bind{x}{t_x},\tcenv_2}{\spred_1}{\spred_2}$ and
      $\imply{\tcenv_1,\bind{a}{k},\tcenv_2}{\spred_1}{\spred_2}$.
      \item (Subst I) If
      $\imply{\tcenv_1,\bind{x}{t_x},\tcenv_2}{\spred_1}{\spred_2}$
      and $\hastype{\tcenv_2}{v_x}{t_x}$, then
      $\imply{\subst{\tcenv_1}{x}{v_x},\tcenv_2}{\subst{\spred_1}{x}{v_x}}
                                             {\subst{\spred_2}{x}{v_x}}$.
      \item (Subst II) If $\imply{\tcenv_1,\bind{a}{k},\tcenv_2}{\spred_1}{\spred_2}$
      and $\isWellFormed{\tcenv_2}{t}{k}$, then
      $\imply{\subst{\tcenv_1}{a}{t},\tcenv_2}{\subst{\spred_1}{a}{t}}
      {\subst{\spred_2}{a}{t}}$.
      \item (Strengthening) If
      $\imply{\bind{y}{\breft{\sbase}{x}{q},\tcenv}}{\spred_1}{\spred_2}$,
      then $\imply{\bind{y}{\sbase},\tcenv}{\csand{\subst{q}{x}{y}}{\spred_1}}
                  {\csand{\subst{q}{x}{y}}{\spred_2}}$.
  \end{enumerate}
\end{requirement}

This axiomatic approach precisely explicates
the properties that are required of
the implication checker in order
to establish the soundness of the
entire refinement type system.
In the future, we can look into
either verifying that these
properties hold for SMT-based
checkers, or even build other
kinds of implication oracles
that adhere to this contract.

\begin{figure}
\judgementHead{Subtyping}{\isSubType{\tcenv}{s}{t}}

\begin{mathpar}   
  \inferrule*[Right=\sFunc]
  { \isSubType{\tcenv}{t_{x2}}{t_{x1}} \quad
    \forall\notmem{y}{\tcenv}. \quad
    \isSubType{\bind{y}{t_{x2}},\tcenv}{\subst{t_1}{x}{y}}{\subst{t_2}{x}{y}} }
  {\isSubType{\tcenv}{\functype{x}{t_{x1}}{t_1}}{\functype{x}{t_{x2}}{t_2}}} \\
\and
  \inferrule*[Right=\sWitn]
  { \hastype{\tcenv}{\sval_x}{\stype_x} \quad
    \isSubType{\tcenv}{\stype}{\subst{\stype'}{x}{\sval_x}}}
  {\isSubType{\tcenv}{\stype}{\existype{x}{\stype_x}{\stype'}}}
\and
  \inferrule*[Right=\sBind]
  { \forall\notmem{y}{\free{\stype}\cup \tcenv}. \quad
    \isSubType{\bind{y}{\stype_x},\tcenv}{\subst{\stype}{x}{y}}{\stype'}
  }
  {\isSubType{\tcenv}{\existype{x}{\stype_x}{\stype}}{\stype'}} \\
\and
  \inferrule*[Right=\sPoly]
  { \forall\notmem{\al'}{\tcenv}. \quad
    \isSubType{\bind{\al'}{\skind},\tcenv}{\subst{\stype_1}{\al}{\al'}}{\subst{\stype_2}{\al}{\al'}} }
  {\isSubType{\tcenv}{\polytype{\al}{\skind}{\stype_1}}{\polytype{\al}{\skind}{\stype_2}}}
\and
  \inferrule*[Right=\sBase]
  {\forall\notmem{y}{\tcenv}. \quad
    \imply{\bind{y}{{\sbase}},\tcenv}{\subst{p_1}{x}{y}}{\subst{p_2}{x}{y}} }
  {\isSubType{\tcenv}{\breft{\sbase}{x}{\spred_1}}{\breft{\sbase}{x}{\spred_2}}}
\end{mathpar}
\caption{Subtyping Rules.}
\label{fig:s}
\label{fig:subtyping}
\label{fig:ent}
\end{figure}

\section{$\sysf$ Soundness}
\label{sec:soundnessF}

Next, we present the metatheory  
of the underlying (unrefined) 
$\sysf$ that, even though it follows the textbook techniques of~\citet{TAPL},
it is a convenient 
stepping stone \emph{towards} 
the metatheory for (refined) $\sysrf$.
In addition, the soundness results 
for $\sysf$ are used \emph{for} 
our full metatheory, as our well-formedness 
judgments require the refinement 
predicate to have the $\sysf$ type 
$\tbool$ thereby avoiding the circularity 
of using a regular typing judgment in the 
antecedents of the well-formedness rules. 
%
%
The \colboth boxes in~\cref{fig:graph}
show the high level outline of the metatheory 
for $\sysf$ which provides a miniaturized model
for $\sysrf$ but without the challenges 
of subtyping
and existentials. Next, we describe the top-level 
type safety result, how it is decomposed into 
progress (\S~\ref{sec:sysf:progress}) and 
preservation (\S~\ref{sec:sysf:preservation}) 
lemmas, and the various technical results that 
support the lemmas.



%
%
The main type safety theorem for 
\sysf states that a well-typed 
term does not get stuck: \ie either 
evaluates to a value or can step 
to another term (progress) 
of the same type (preservation).
The judgment \hasftype{\tcenv}{\sexpr}{\sftype} 
is defined in~\cref{fig:typing} without the 
grey boxes, and for clarity we use $\sftype$ for \sysf types. 

\begin{theorem} (Type Safety)
If $\hasftype{\varnothing}{\sexpr}{\sftype}$ and $\evalsTo{\sexpr}{\sexpr'}$,
then $\sexpr'$ is a value or $\sexpr' \step \sexpr''$ 
for some $\sexpr''$. 
\end{theorem}
We prove type safety by induction on the 
length of the sequence of steps comprising 
$\evalsTo{\sexpr}{\sexpr'}$, using the 
preservation and progress lemmas.

\subsection{Progress} \label{sec:sysf:progress}
The progress lemma says that a well-typed term is either a value 
or steps to some other term.
\begin{lemma} (Progress) \label{lem:progressF} 
If $\hasftype{\varnothing}{\sexpr}{\sftype}$, 
then $\sexpr$ is a value or $\sexpr \step \sexpr'$ for some $\sexpr'$.
\end{lemma}
Proof of progress requires a \emph{Canonical Forms} 
lemma (\cref{lem:canonicalF}) which describes the 
shape of well-typed values and some key properties 
about the built-in \emph{Primitives} (\cref{lem:primitivesF}).
We also implicitly use an \emph{Inversion of Typing} 
lemma (\cref{lem:inversionF}) which describes the shape of the 
type of well-typed terms and its subterms. For $\sysf$, 
unlike $\sysrf$, this lemma
is trivial because the typing relation is syntax-directed.

%

\begin{lemma}\label{lem:canonicalF} (Canonical Forms) 
\begin{enumerate}
    \item If $\hasftype{\varnothing}{v}{\tbool}$, 
        then $v = \ttrue$ or $v = \tfalse$.
    \item If $\hasftype{\varnothing}{v}{\tint}$, then $v$ is an integer constant.
    \item If $\hasftype{\varnothing}{v}{\funcftype{\sftype}{\sftype'}}$, 
        then either $v = \vabs{x}{\sexpr}$ or $v = \sconst$, 
        a constant function where 
        $\sconst \in \{\wedge, \vee, \neg, \leftrightarrow \}$.
    \item If $\hasftype{\varnothing}{v}{\polytype{\al}{\skind}{\sftype}}$, 
        then either $v = \tabs{\al}{\skind}{\sexpr}$ 
        or $v = \sconst$, a polymorphic constant $\sconst \in \{\leq, =\}$.
    \item If $\isWFFT{\varnothing}{\sftype}{\skbase}$,
        then $\sftype = \tbool$ or $\sftype = \tint$.
\end{enumerate}
\end{lemma}

\begin{lemma}\label{lem:inversionF} (Inversion of Typing) 
    \begin{enumerate}
        \item If $\hasftype{\tcenv}{c}{\sftype}$, 
            then $\sftype = \forgetreft{\ty{c}}$.
        \item If $\hasftype{\tcenv}{x}{\sftype}$, 
            then $\bind{x}{\sftype} \in \tcenv$.
        \item If $\hasftype{\tcenv}{\app{e}{e_x}}{\sftype}$,
            then there is some type $\sftype_x$ such that  
            $\hasftype{\tcenv}{e}{\funcftype{\sftype_x}{\sftype}}$ and
            $\hasftype{\tcenv}{e_x}{\sftype_x}$.
        \item If $\hasftype{\tcenv}{\vabs{x}{e}}{\sftype}$,\! 
            then $\sftype = \funcftype{\sftype_x}{\sftype'}$ and
            $\hasftype{\bind{y}{\sftype_x},\tcenv}{\subst{e}{x}{y}}{\sftype'}$
            for any $\notmem{y}{\tcenv}$ and well-formed $\sftype_x$.
        \item If $\hasftype{\tcenv}{\tyapp{e}{t}}{\sftype}$,\! then there is some
            type $\sigma$ and kind $\skind$ such that 
            $\hasftype{\tcenv}{e}{\polytype{\al}{\skind}{\sigma}}$
            and $\sftype = \subst{\sigma}{\al}{\forgetreft{t}}$.
        \item If $\hasftype{\tcenv}{\tabs{\al}{\skind}{e}}{\sftype}$, then
            there is some type $\sftype'$ and kind $\skind$ such that
            $\sftype = {\polytype{\al}{\skind}{\sftype'}}$ and
            $\hasftype{\bind{\al'}{\skind},\tcenv}{\subst{e}{\al}{\al'}}
            {\subst{\sftype'}{\al}{\al'}}$ for some $\notmem{\al'}{\tcenv}$.
        \item If $\hasftype{\tcenv}{\eletin{x}{e_x}{e}}{\sftype}$, then
            there is some type $\sftype_x$ and $\notmem{y}{\tcenv}$ such that
            $\hasftype{\tcenv}{e_x}{\sftype_x}$ and 
            $\hasftype{\bind{y}{\sftype_x},\tcenv}{\subst{e}{x}{y}}{\sftype}$.
        \item If $\hasftype{\tcenv}{\tyann{e}{t}}{\sftype}$, then 
            $\sftype = \forgetreft{t}$ and $\hasftype{\tcenv}{e}{\sftype}$.
    \end{enumerate}
\end{lemma}

%

\begin{lemma}\label{lem:primitivesF}(Primitives) 
For each built-in primitive $c$, 
\begin{enumerate} 
\item If $\forgetreft{\ty{\sconst}} = \funcftype{\sftype_x}{\sftype}$
    and $\hasftype{\varnothing}{v_x}{\sftype_x}$, 
    then 
    $\hasftype{\varnothing}{\delta(\sconst,v_x)}{\sftype}$.
\item If $\forgetreft{\ty{\sconst}} = \polytype{\al}{\skind}{\sftype}$ 
    and $\isWellFormed{\varnothing}{\sftype_\al}{\skind}$, 
    then 
    $\hasftype{\varnothing}{\delta_T(\sconst,\sftype_\al)}{\subst{\sftype}{\al}{\sftype_\al}}$.
    \end{enumerate}
\end{lemma}

Lemmas~\ref{lem:canonicalF} and~\ref{lem:inversionF} are proved without
induction by inspection of the derivation tree, 
while lemma~\ref{lem:primitivesF} relies on 
the Primitives Requirement~\ref{lem:prim-typing}. 

\subsection{Preservation} \label{sec:sysf:preservation}
The preservation lemma states that $\sysf$ typing is preserved
by evaluation.
\begin{lemma} (Preservation) \label{lem:preservationF} 
If $\hasftype{\varnothing}{\sexpr}{\sftype}$ and $\sexpr \step \sexpr'$, 
then $\hasftype{\varnothing}{\sexpr'}{\sftype}$.
\end{lemma}    

The proof is by structural induction on the 
derivation of the typing judgment. 
We use the determinism of the operational 
semantics (Lemma~\ref{lem:step-determ}) and 
the canonical forms lemma to case split 
on $\sexpr$ to determine $\sexpr'$.
The interesting cases are for \fApp and \fTApp.
For applications of primitives, preservation 
requires the Primitives~\Cref{lem:primitivesF},
while the general case needs a Substitution~\Cref{lem:substitutionF}.

\mypara{Substitution Lemma}
To prove type preservation when
a lambda or type abstraction is applied, 
we proved that the substituted result 
has the same type, as established by 
the substitution lemma:
\begin{lemma}(Substitution)\label{lem:substitutionF}
If $\hasftype{\tcenv}{\sval_x}{\sftype_x}$ 
and $\isWFFT{\tcenv}{\forgetreft{\stype_{\al}}}{k_{\al}}$, then 
\begin{enumerate}
\item if\; $\hasftype{\tcenv', \bind{x}{\sftype_x}, \tcenv}{\sexpr}{\sftype}$
    and\; $\isWFFE{\tcenv}$, then
    ${\hasftype{\tcenv', \tcenv}{\subst{\sexpr}{x}{\sval_x}}{\sftype}}$.
\item if\; $\hasftype{\tcenv', \bind{\al}{k_{\al}}, \tcenv}{\sexpr}{\sftype}$
    and\; $\isWFFE{\tcenv}$, then
    ${\hasftype{\subst{\tcenv'}{\al}{\forgetreft{\stype_{\al}}}, \tcenv}
               {\subst{\sexpr}{\al}{\stype_{\al}}}
               {\subst{\sftype}{\al}{\forgetreft{\stype_{\al}}}}}$.
\end{enumerate}
\end{lemma}

The proof goes by induction on the derivation tree. 
Because we encoded our typing rules using 
cofinite quantification (\S~\ref{sec:lang:static})
the proof does not require a renaming lemma, but 
the rules that lookup environments 
(rules \tVar and \wtVar) do need a lemma the below \emph{Weakening} 
\Cref{lem:weakeningF}. 

\begin{lemma}\label{lem:weakeningF}(Weakening Environments) 
If\; $\hasftype{\tcenv_1, \tcenv_2}{\sexpr}{\sftype}$ and 
   $x,\al \not\in {\tcenv_1, \tcenv_2}$,
then 
\begin{enumerate} 
    \item 
    \hasftype{\tcenv_1, \bind{x}{\sftype_x}, \tcenv_2}{\sexpr}{\sftype}
    \ \text{and}\ \ 
    (2)\ \hasftype{\tcenv_1, \bind{\al}{\skind}, \tcenv_2}{\sexpr}{\sftype}.
\end{enumerate}
\end{lemma}

\section{$\sysrf$ Soundness} \label{sec:soundness}

We proceed to the metatheory
of $\sysrf$ by fleshing out
the skeleton of \colboth lemmas
in~\cref{fig:graph} (which
have similar statements to the $\sysf$
versions) and describing the three
regions (\S~\ref{sec:overview:rf})
needed to establish the properties
of inversion, substitution, and narrowing.


\mypara{Type Safety}
The top-level type safety theorem, like \sysf,
combines progress and preservation
\begin{theorem} (Type Safety) \label{lem:soundness}
    If $\hastype{\varnothing}{\sexpr}{\stype}$ and $\evalsTo{\sexpr}{\sexpr'}$,
    then $\sexpr'$ is a value or $\sexpr' \step \sexpr''$
    for some $\sexpr''$.
\end{theorem}
\begin{lemma} (Progress) \label{lem:progress}
If $\hastype{\varnothing}{\sexpr}{\stype}$,
then $\sexpr$ is a value or $\sexpr \step \sexpr'$ for some $\sexpr'$.
\end{lemma}
\begin{lemma} (Preservation) \label{lem:preservation}
    If $\hastype{\varnothing}{\sexpr}{\stype}$ and $\sexpr \step \sexpr'$,
    then $\hastype{\varnothing}{\sexpr'}{\stype}$.
\end{lemma}
Next, let's see the three main ways in
which the proof of~\cref{lem:progress}
differs from $\sysf$.

\subsection{Inversion of Typing Judgments}
\label{sec:soundness:inversion}

The \colinversion region of~\cref{fig:graph}
accounts for the fact that, due to subtyping
chains, the typing judgment in \sysrf is not
syntax-directed.
%
%
First, we establish that subtyping is transitive

\begin{lemma} (Transitivity)
    If\; $\isWellFormed{\tcenv}{t_1\!}{\! k_1}$,
       $\isWellFormed{\tcenv}{t_3\!}{\! k_3}$,
       $\isWellFormedE{\tcenv}$,
       $\isSubType{\tcenv}{t_1}{t_2}$, 
       $\isSubType{\tcenv}{t_2}{t_3}$, then
       $\isSubType{\tcenv}{t_1}{t_3}$.
\end{lemma}
The proof consists of a case-split
on the possible rules for $\isSubType{\tcenv}{t_1}{t_2}$
and $\isSubType{\tcenv}{t_2}{t_3}$.
When the last rule used
in the former is \sWitn and the
latter is \sBind, we require the
Substitution~\Cref{lem:subst}.
As \citet{Aydemir05},
we use the Narrowing~\Cref{lem:narrowing}
for the transitivity for function types.

\mypara{Inverting Typing Judgments}
We use the transitivity of subtyping to prove
some non-trivial lemmas that let us ``invert''
the typing judgments to recover information
about the underlying terms and types.
We describe the non-trivial case which
pertains to type and value abstractions:

\begin{lemma} (Inversion of $\tAbs$, $\tTAbs$)
    \begin{enumerate}
    \item If $\hastype{\tcenv}{(\vabs{w}{\sexpr})}{\functype{x}{\stype_x}{\stype}}$
    and $\isWellFormedE{\tcenv}$,
    then for all $\notmem{y}{{\tcenv}}$
    we have $\hastype{\bind{y}{\stype_x},\tcenv}{\subst{\sexpr}{w}{y}}{\subst{\stype}{x}{y}}$.
    \item If $\hastype{\tcenv}{(\tabs{\al_1}{\skind_1}{\sexpr})}{\polytype{\al}{\skind}{\stype}}$
    and $\isWellFormedE{\tcenv}$,
    then for every $\notmem{\al'}{{\tcenv}}$
    we have $\hastype{\bind{\al'}{\skind},\tcenv}{\subst{\sexpr}{\al_1}{\al'}}{\subst{\stype}{\al}{\al'}}$. 
    \end{enumerate}
\end{lemma}
If $\hastype{\tcenv}{(\vabs{w}{e})}{\functype{x}{t_x}{t}}$,
then we cannot directly invert the typing judgment
to get a typing for the body $e$ of $\vabs{w}{e}$.
Perhaps the last rule used was \tSub,
and inversion only tells us that there
exists a type $t_1$ such that
$\hastype{\tcenv}{(\vabs{w}{e})}{t_1}$
and $\isSubType{\tcenv}{t_1}{\functype{x}{t_x}{t}}$.
Inverting again, we may in fact find a chain
of types $t_{i+1} \subt t_i \subt \cdots \subt t_2 \subt t_1$
which can be arbitrarily long.
But the proof tree must be finite so eventually
we find a type $\functype{w}{s_w}{s}$ such that
$\hastype{\tcenv}{(\vabs{w}{e})}{\functype{w}{s_w}{s}}$
and $\isSubType{\tcenv}{\functype{w}{s_w}{s}}{\functype{x}{t_x}{t}}$
(by transitivity)
and the last rule was $\tAbs$.
Then inversion gives us that for any $\notmem{y}{\tcenv}$
we have
$\hastype{\bind{y}{s_w}, \tcenv}{e}{\subst{s}{w}{y}}$.
To get the desired typing judgment, we must
use the Narrowing~\Cref{lem:narrowing} to obtain
$\hastype{\bind{y}{t_x}, \tcenv}{e}{\subst{s}{w}{y}}$.
Finally, we use $\tSub$ to derive
$\hastype{\bind{y}{t_x}, \tcenv}{e}{\subst{t}{w}{y}}$.

\subsection{Substitution Lemma}

The main result in the \colsubtyping
region of~\cref{fig:graph} is
the Substitution Lemma.
The biggest difference between the
$\sysf$ and $\sysrf$ metatheories
is the introduction of a mutual
dependency between the lemmas
for typing and subtyping judgments.
Due to this dependency, the substitution
lemma, and the weakening lemma
on which it depends must now be proven
in a mutually recursive form for both
typing and subtyping judgments:

\begin{lemma}\label{substitution} \label{lem:subst} (Substitution)
    \begin{itemize}
        \item If $\isSubType{\tcenv_1,\bind{x}{t_x},\tcenv_2}{s}{t}$,
                 $\isWellFormedE{\tcenv_2}$,
            and $\hastype{\tcenv_2}{v_x}{t_x}$,
            then $\isSubType{\subst{\tcenv_1}{x}{v_x},\tcenv_2}{\subst{s}{x}{v_x}}{\subst{t}{x}{v_x}}$.
        \item If $\hastype{\tcenv_1,\bind{x}{t_x},\tcenv_2}{e}{t}$,
                 $\isWellFormedE{\tcenv_2}$,
            and $\hastype{\tcenv_2}{v_x}{t_x}$,
            then $\hastype{\subst{\tcenv_1}{x}{v_x},\tcenv_2}{\subst{e}{x}{v_x}}{\subst{t}{x}{v_x}}$.
        \item If $\isSubType{\tcenv_1,\bind{\al}{\skind},\tcenv_2}{s}{t}$,
                $\isWellFormedE{\tcenv_2}$,
            and $\isWellFormed{\tcenv_2}{t_\al}{k}$,
            then $\isSubType{\subst{\tcenv_1}{\al}{t_\al},\tcenv_2}{\subst{s}{\al}{t_\al}}{\subst{t}{\al}{t_\al}}$.
        \item If $\hastype{\tcenv_1,\bind{\al}{\skind},\tcenv_2}{e}{t}$,
                 $\isWellFormedE{\tcenv_2}$,
            and $\isWellFormed{\tcenv_2}{t_\al}{\skind}$,
            then $\hastype{\subst{\tcenv_1}{\al}{t_\al},\tcenv_2}{\subst{e}{\al}{t_\al}}{\subst{t}{\al}{t_\al}}$.
    \end{itemize}
\end{lemma}

The main difficulty arises
in substituting some type
$t_\al$ for variable $\al$
in $\isSubType{\tcenv_1,\bind{\al}{\skind},\tcenv_2}{\breft{\al}{x_1}{p}}{\breft{\al}{x_2}{q}}$
because we must deal with
strengthening $t_\al$ by
the refinements $p$ and
$q$ respectively.
As with the $\sysf$ metatheory, the proof of the substitution lemma
does not require renaming, but does
require a lemmas that let us \emph{weaken}
environments (\cref{lem:weakening})
in typing and subtyping judgments.

\begin{lemma}\label{lem:weakening} (Weakening Environments)
If $\notmem{x,\al}{{\tcenv_1, \tcenv_2}}$, then
    \begin{enumerate}
        \item if $\hastype{\tcenv_1, \tcenv_2}{\sexpr}{\stype}$
        then $\hastype{\tcenv_1, \bind{x}{\stype_x}, \tcenv_2}{\sexpr}{\stype}$ and $\hastype{\tcenv_1, \bind{\al}{\skind}, \tcenv_2}{\sexpr}{\stype}$.
        \item if $\isSubType{\tcenv_1, \tcenv_2}{s}{t}$
        then $\isSubType{\tcenv_1, \bind{x}{\stype_x}, \tcenv_2}{s}{t}$
        and $\isSubType{\tcenv_1, \bind{\al}{\skind}, \tcenv_2}{s}{t}$.
    \end{enumerate}
\end{lemma}

The proof is by mutual induction
on the derivation of the typing and subtyping judgments.

\subsection{Narrowing} \label{sec:narrowing}

The narrowing lemma says that whenever
we have a judgment where a binding
$x\bindt t_x$ appears in the binding
environment, we can replace $t_x$
by any subtype $s_x$.
The intuition here is that the judgment
holds under the replacement because we
are making the context more specific.

\begin{lemma} \label{subtype-env} \label{lem:narrowing} (Narrowing)
    If $\tcenv_2 \vdash s_x <: t_x$,  $\isWellFormed{\tcenv_2}{s_x}{k_x}$,
    and $\isWellFormedE{\tcenv_2}$ then
    \begin{enumerate}
        \item if $\tcenv_1, \bind{x}{t_x} \tcenv_2 \vdash_w t : k$, then
                 $\tcenv_1, x\bindt s_x, \tcenv_2 \vdash_w t : k $.
        \item if $\tcenv_1, \bind{x}{t_x}, \tcenv_2 \vdash t_1 <: t_2$, then
                 $\tcenv_1, x\bindt s_x, \tcenv_2 \vdash t_1 <: t_2$.
        \item if $\tcenv_1, \bind{x}{t_x}, \tcenv_2 \vdash e : t$, then
                 $\tcenv_1, x\bindt s_x, \tcenv_2 \vdash e : t$.
    \end{enumerate}
\end{lemma}

The narrowing proof
requires an Exact Typing~\Cref{lem:exact}
which says that a subtyping
judgment $\isSubType{\tcenv}{s}{t}$
is preserved after
selfification on both types.
Similarly whenever we can type
a value $v$ at type $t$ then we
also type $v$ at the type $t$
selfified by $v$.

\begin{lemma} (Exact Typing) \label{lem:exact}
\begin{enumerate}
  \item If $\hastype{\tcenv}{e}{t}$, $\isWellFormedE{\tcenv}$, $\isWellFormed{\tcenv}{t}{k}$, and $\isSubType{\tcenv}{s}{t}$, then $\isSubType{\tcenv}{{\rm self}(s, v, k)}{{\rm self}(t, v, k)}$.
  \item If $\hastype{\tcenv}{v}{t}$, $\isWellFormedE{\tcenv}$, and $\isWellFormed{\tcenv}{t}{k}$, then $\hastype{\tcenv}{v}{{\rm self}(t, v, k)}$. 
\end{enumerate}
\end{lemma}

\section{Refined Data Propositions}
\label{sec:data-props}

In \S~\ref{sec:implementation} we will present how 
type soundness \sysrf is encoded in \lh. 
Here we present \textit{refined data propositions}, 
a novel feature of \lh that made such 
a meta-theoretic development possible. 
Intuitively, refined data propositions encode 
\coq-style inductive predicates to permit 
constructive reasoning about potentially non-terminating 
properties, as required for meta-theoretic developments. 
(\citet{lweb} developed a meta-theoretic proof in \lh 
without refined data propositions, but assumed a terminating 
evaluation relation; see~\S~\ref{sec:related}.)

Refined data propositions
encode inductive predicates in \lh by refining Haskell's 
data types, allowing the programmer to write plain
Haskell functions to provide constructive proofs for user-defined
propositions.
Here, for exposition, we present the four steps we followed  
in the mechanization of \sysrf to 
define the ``hastype'' proposition
and then use it to type the primitive one. 

\mypara{Step 1: Reifying Propositions as Data}
Our first step is to represent the different propositions
of interest as plain Haskell data.
For example, we can define the following types 
(with the suffix \ha{Pr} for ``proposition''):
\begin{mcode}
    data HasTyPr   = HasTy    Env  Expr Type
    data IsSubTyPr = IsSubTy  Env  Type Type
\end{mcode}
Thus, \ha{HasTy $\gamma$ e t} and \ha{IsSubTy $\gamma$ t t'}
respectively represent the \emph{propositions} 
that \ha{e} can be typed as \ha{t} under environment $\gamma$ and
that \ha{t} is a subtype of \ha{t'} under $\gamma$.

\mypara{Step 2: Reifying Evidence as Data}
Next, we reify evidence, \ie \emph{derivation trees},
also as data by defining Haskell data types with a
\emph{single constructor per derivation rule}.
For example, we define the Haskell data type \ha{HasTyEv}
to encode the typing rules of~\cref{fig:t}, using a single
constructor, whose name matches the corresponding rule:
\begin{mcode}
  data HasTyEv where
    TPrim :: Env -> Prim -> HasTyEv
    TSub  :: Env -> Expr -> Type -> Type -> HasTyEv -> IsSubTyEv -> HasTyEv
    ...
\end{mcode}
Using these data one can construct derivation trees. 
For instance,  
\ha{TPrim Empty (PInt 1) :: HasTyEv} is the tree that types 
the primitive one under the empty environment. 

\mypara{Step 3: Relating Evidence to its Propositions}
Next, we specify the relationship between the evidence and the proposition
that it establishes, via a refinement-level \emph{uninterpreted function}
that maps evidence to its proposition:
\begin{mcode}
    measure hasTyEvPr   :: HasTyEv -> HasTyPr
    measure isSubTyEvPr :: IsSubTyEv -> IsSubTyPr
\end{mcode}
The above signatures declare that \ha{hasTyEvPr} (resp. \ha{isSubTyEvPr})
is a refinement-level function that maps has-type (resp. is-subtype)
evidence to its corresponding proposition.
We can now use these uninterpreted functions to define \emph{type aliases}
that denote well-formed evidence that establishes a proposition.
For example, consider the (refined) type aliases
\begin{mcode}
  type HasTy   $\gamma$ e t  = {ev:HasTyEv   | hasTyEvPr ev == HasTyPr $\gamma$ e t }
  type IsSubTy $\gamma$ t t' = {ev:IsSubTyEv | isSubTyEvPr ev == IsSubTyPr $\gamma$ t t' }
\end{mcode}
The definition stipulates that the type \ha{HasTy $\gamma$ e t}
is inhabited by evidence (of type \ha{HasTyEv}) that
establishes the typing proposition \ha{HasTyPr $\gamma$ e t}.
Similarly \ha{IsSubTy $\gamma$ t t'} is inhabited by evidence
(of type \ha{IsSubTyEv}) that establishes the sub-typing
proposition \ha{IsSubTyPr $\gamma$ t t'}.
Note that the first three steps have only defined separate data types
for propositions and evidence, and \emph{specified} the relationship
between them via uninterpreted functions in the refinement logic.

\mypara{Step 4: Refining Evidence to Establish Propositions}
Our final step is to \emph{implement} the relationship between
evidence and propositions \emph{refining} the types of the evidence
data constructors (rules) with input types (pre-conditions) that require
the rules' premises hold, and output types (post-conditions) that ensure
the rules' conclusions.
For example, we connect the evidence and proposition for the
typing relation by refining the data constructors for \ha{HasTyEv}
using the premises and conclusions of their respecting typing rule
from ~\cref{fig:t}.
\begin{mcode}
  data HasTyEv where
    TPrim :: $\gamma$:Env -> c:Prim ->
             HasTy $\gamma$ (Prim c) (ty c)
    TSub :: $\gamma$:Env -> e:Expr -> t:Type -> t':Type ->
            HasTy $\gamma$ e t -> IsSubTy $\gamma$ t t' ->
            HasTy $\gamma$ e t'
    ...
\end{mcode}
The constructors \ha{TPrim} and \ha{TSub} respectively
encode the rules \tPrim and \tSub rules (with the latter
being simplified here to elide the well-formedness of \ha{t'}).
The refinements on the input types, 
which encode the premises of the rules,
are checked whenever these constructors are used. 
The refinement on the 
output type (being evidence of a specific proposition)
is axiomatized to encode the conclusion of the rules.
Consider the type for \ha{TSub}.
The first line says ``for all $\gamma, e, t, t'$'',
the second line (premise) says, ``given evidence that $\hastype{\gamma}{e}{t}$ and $\isSubType{\gamma}{t}{t'}$'',
and the last line (conclusion) says the constructor returns ``evidence that $\hastype{\gamma}{e}{t'}$''.

\mypara{Programs as Constructive Proofs}
Thus, the constructor refinements crucially ensure that only well-formed pieces of evidence
can be constructed, and simultaneously, precisely track the proposition established
by the evidence.
This lets the programmer write plain Haskell terms as constructive proofs, and \lh ensures
that those terms indeed establish the proposition stipulated by their type.
For example, the below Haskell term is proof that the literal \ha{1} has the type
$\breft{\Int}{\vv}{\vv = 1}$
\begin{mcode}
  oneTy :: HasTy Empty (EPrim (PInt 1)) {v:Int | v == 1}
  oneTy = TPrim Empty (PInt 1)
\end{mcode}
If instead, the programmer modified their code to
\ha{oneTy = TPrim Empty (PInt 2)}, \lh would reject
the program as the modified evidence does not establish
the proposition described in the type.
%


\section{Mechanization}
\label{sec:implementation}

We have mechanized soundness of \sysrf in \lh.
The mechanization, submitted as anonymous supplementary
material, uses refined data propositions
(\S~\ref{sec:data-props}) to specify the
static (\eg typing, subtyping, well-formedness)
and dynamic (\ie small-step transitions and their closure)
semantics of \sysrf.
The mechanization is simplified by SMT-automation (\S~\ref{impl:settheory}),
uses a co-finite encoding for reasoning about variables (\S~\ref{subsec:implementation:co-finite}),
and comprises of proofs implemented as recursive functions that
constructively manipulate the evidence to establish propositions by
induction (\S~\ref{impl:proofs}).
Note that while Haskell types are inhabited
by diverging $\bot$ values, \lh's totality,
termination and type checks ensure that all
cases are handled, the induction (recursion)
is well-founded, and that the proofs (programs)
indeed establish the given propositions (types).

\subsection{SMT Solvers and Set Theory} \label{impl:settheory}

The most tedious part in mechanization of metatheories
is the establishment of invariants about variables,
for example uniqueness and freshness.
\lh offers a built-in, SMT automated support for the
theory of sets, which simplifies establishing such
invariants.

\mypara{Set of Free Variables}
Our proof mechanization defines the Haskell function
\ha{fv} that returns the \ha{Set} of free variable names
that appear in its argument.
\begin{mcode}
  measure fv
  fv :: Expr -> S.Set VName
  fv (EVar x)    = S.singleton x
  fv (ELam e)    = fv e
  fv (EApp e e') = S.union (fv e) (fv e')
  ... -- other cases
\end{mcode}
In the above (incomplete) definition, \ha{S}
is used to qualify the standard \verb+Data.Set+ Haskell library.
\lh embeds the functions of \verb+Data.Set+ to SMT
set operators (encoded as a map to booleans).
For example, \ha{S.union} is treated as the
logical set union operator $\cup$.
Further, we lift \ha{fv} into the refinement
logic using the \ha{measure fv} annotation.
%
%
The measure definition  defines the logical
function \ha{fv} in the logic in a way that
lets the SMT solver reason about the semantics
of \ha{fv} in a \textit{decidable} fashion,
as an uninterpreted function refining the
type of each \ha{Expr} data constructor
\cite{sprite}.
This embedding, combined with the SMT solver's
support for the theory of sets, lets \lh prove
properties about expressions' free variables
``for free''.

\mypara{Intrinsic Verification}
For example, consider the function
\ha{subFV x vx e} which substitutes
the free variable \ha{x} with
\ha{vx} in \ha{e}.
The refinement type of \ha{subFV} 
describes the free variables of the result.
\begin{mcode}
  subFV :: x:VName -> vx:{Expr | isVal vx } -> e:Expr
        -> {e':Expr | fv e' $\subseteq$ (fv vx $\cup$ (fv e \ x)) && (isVal e => isVal e')}
  subFV x vx (EVar y)    = if x == y then vx else EVar y
  subFV x vx (ELam   e)  = ELam (subFV x vx e)
  subFV x vx (EApp e e') = EApp (subFV x vx e) (subFV x vx e')
  ... -- other cases
\end{mcode}
The refinement type post condition specifies
that the free variables after substitution is a subset
of the free variables on the two argument expressions,
excluding \ha{x}, \ie
$\mathtt{fv}(\subst{e}{x}{v_x}) \subseteq
\mathtt{fv}(e) \cup (\mathtt{fv}(v_x) \setminus \{x\})$.
This specification is proved \emph{intrinsically},
that is the definition of \ha{subFV} constitutes the proof
(no user aid is required) and, importantly,
the specification is automatically established each time
the function \ha{subFV} is called.
So, the user does not have to provide explicit hints to reason
about free variables of substituted expressions.

The specification of \ha{subFV} shows another example
of SMT-based proof simplification: the Haskell boolean
function \ha{isVal} is another \ha{measure} that defines
when an expression is a \emph{value}. By doing so, we can
intrinsically prove that that value property is preserved
by substitution, as stated by the second (implication) conjunct
in the output of \ha{subFV}.

\mypara{Freshness}
\lh's support for sets simplifies
defining a \ha{fresh} function,
which is often challenging\footnote{For example, \coq cannot fold over a set,
and so a more complex combination of tactics is
required to generate a fresh name.}.
\ha{fresh xs} returns a variable that provably
does not belong to its input \ha{xs}.
\begin{mcode}
  fresh :: xs:S.Set VName -> { x:VName | x $\not\in$ xs }
  fresh xs = n ? above_max n xs'
    where n    = 1 + maxs xs'
          xs'  = S.fromList xs

  maxs :: [VName] -> VName
  maxs []     = 0
  maxs (x:xs) = if maxs xs < x then x else maxs xs

  above_max :: x:VName -> xs:{[VName]|maxs xs < x} -> {x $\not\in$ elems xs}
  above_max _ []     = ()
  above_max x (_:xs) = above_max x xs
\end{mcode}
The \ha{fresh} function returns \ha{n}: the maximum element of the set
increased by one.
To compute the maximum element we convert the set to a list
and use the inductively defined \ha{maxs} functions.
To prove \ha{fresh}'s intrinsic specification we use
an extrinsic, \ie explicit, lemma \ha{above_max n xs'}
that, via the \ha{(?)} combinator of type \ha{a ->  b ->  a},
tells \lh that \ha{n} is not in the set \ha{xs}.
This extrinsic lemma is itself trivially proved
by induction on \ha{xs} and SMT automation.

\subsection{Co-finite Quantification}
\label{subsec:implementation:co-finite}
\begin{figure}
\begin{mcode}
-- Standard Existential Rule
TAbsEx  :: $\gamma$:Env -> t$_x$:Type -> e:Expr -> t:Type
        -> y:{VName | y $\not\in$ dom $\gamma$ }
        -> HasTy ((y,t$_x$):$\gamma$) (unbind y e) (unbindT y t)
        -> HasTy $\gamma$ (ELam e) (TFunc t$_x$ t)

-- Co-finitely Quantified Rule
TAbs    :: $\gamma$:Env -> t$_x$:Type -> e:Expr -> t:Type -> l:S.Set VName
        -> (y:{VName|y $\not\in$ l} -> HasTy ((y,t$_x$):$\gamma$) (unbind y e) (unbindT y t))
        -> HasTy $\gamma$ (ELam e) (TFunc t$_x$ t)

-- Note: All rules also include k$_{\color{gray_ulisses}x}$:Kind -> WfType ${\color{gray_ulisses}\gamma}$ t$_{\color{gray_ulisses}x}$ k$_{\color{gray_ulisses}x}$ elided for clarity.
\end{mcode}
\caption{Encoding of Co-finitely Quantified Rules.}
\label{fig:impl:co-finite}
\end{figure}

%

To encode the rules that need a fresh free variable name
we use the co-finite quantification of~\citet{AydemirCPPW08}, as discussed
in~\S~\ref{sec:lang:static}.
\Cref{fig:impl:co-finite} presents
this encoding using the \tAbs rule
as an example.
The standard abstraction rule
(rule \tAbsEx in~\S~\ref{sec:lang:static})
requires to provide a concrete fresh name,
which is encoded in the second line of
\ha{TAbsEx} as the \ha{y:$\{$VName | y $\not\in$ dom $\gamma\}$} argument.
The co-finitely quantified encoding of the rule \ha{TAbs}, instead,
states that there exists a specified finite set of excluded names, namely \ha{l},
and requires that the sub-derivation holds for any name \ha{y}
that does not belong in \ha{l}.
That is, the premise is turned into a function
that, given the name \ha{y}, returns the sub-derivation.
This encoding greatly simplifies our mechanization,
since the premises are no more tied to concrete names,
eliminating the need for renaming lemmas.
We will often take \ha{l} to be the domain
of the environment, but the ability to choose \ha{l}
gives us the flexibility when constructing derivations
to exclude additional names that clash
with another part of a proof.

%
%
%
%

\subsection{Inductive Proofs as Recursive Functions}
\label{impl:proofs}

The majority of our proofs are by induction on derivations.
These proofs are written as recursive Haskell functions that
operate over the refined data types reifying the respective
derivations. \lh ensures the proofs are valid by checking that
they are inductive (\ie the recursion is well-founded),
handle all cases (\ie the function is total) and
establish the desired properties (\ie witnesses the appropriate proposition.)

\mypara{Preservation (Theorem~\ref{lem:preservation})}
relates the \ha{HasTy} data proposition of~\S~\ref{sec:data-props}
with a \ha{Step} data proposition that encodes~\cref{fig:opsem}
and is proved by induction on the type derivation tree.
The subtyping case requires an induction
while the primitive case is impossible (\cref{lem:step-determ}):
\begin{mcode}
  preservation :: e:Expr -> t:Type -> e':Expr -> HasTy Empty e t -> Step e e'
               -> HasTy Empty e' t

  preservation _e _t e' (TSub Empty e t' t e_has_t' t'_sub_t) e_step_e'
    = TSub Empty e' t' t e'_has_t' t'_sub_t
        where e'_has_t' = preservation e t' e' e_has_t' e_step_e'
  preservation e _ e' (TPrim _ _) step
    = impossible "value" ? lemValStep e e' step -- $e \step e' \Rightarrow \neg (\text{isVal}\ e)$
  preservation e _ e' (TAbs {}) step
    = impossible "value" ? lemValStep e e' step -- $e \step e' \Rightarrow \neg (\text{isVal}\ e)$
  ...

  impossible :: {v:String | false} -> a
  lemValStep :: e:Expr -> e':Expr -> Step e e' -> {$\neg$(isVal e)}
\end{mcode}

In the \ha{TSub} case we note that \lh knows that
the expression argument \ha{_e} is equal to the
the subtyping parameter \ha{e}.
Further, the termination checker will ensure
that the inductive call happens on the smaller
derivation subtree.
The \ha{TPrim} case goes by contradiction since
primitives cannot step.
We separately proved that values cannot step
in the \ha{lemValStep} lemma, which here
is combined with the fact that \ha{e} is a value
to allow the call of the false-precondition \ha{impossible}.
Finally, \lh's totality checker ensures
all the cases of \ha{HasTypEv} are covered,
and the termination checker ensures
the proof is well-founded. 

\mypara{Progress (Theorem~\ref{lem:progress})}
ensures that a well-typed expression is a value
\textit{or} there exists an expression to which
it steps.
To express this claim we used Haskell's \ha{Either}
to encode disjunction that contain pairs (refined to be dependent)
to encode existentials.
%
%
\begin{mcode}
  progress :: e:Expr -> t:Type -> HasTy Empty e t 
           -> Either {isVal e}  (e'::Expr, Step e e')

  progress _ _ (TSub Empty e t' t e_has_t' _) = progress e t' e_has_t'
  progress _ _ (TPrim _ _)                    = Left ()
  progress _ _ (TAbs {})                      = Left ()
  ...
\end{mcode}
The proofs of the \ha{TSub} and \ha{TPrim}
cases are easily done by, respectively,
an inductive call and establishing is-Value.
The more interesting cases require us to case-split
on the inductive call in order to get access
to the existential witness.

\mypara{Soundness (Theorem~\ref{lem:soundness})}
ensures that a well-typed expression will not get stuck,
that is, it will either reach a value or keep evaluating.
We reify evaluation sequences with a refined data
proposition \ha{Steps e$_0$ e} with a reflexive
and a transitive (recursive) constructor.
Our soundness proof goes by induction on the
evaluation sequence.
\begin{mcode}
  soundness :: e$_0$:Expr -> t:Type -> e:Expr -> HasTy Empty e$_0$ t -> Steps e$_0$ e
            -> Either {isVal e}  (e$_i$::Expr, Step e e$_i$)

  soundness _e$_0$ t _e e$_0$_has_t e$_0$_evals_e = case e$_0$_evals_e of
     Refl e$_0$ -> progress e$_0$ t e$_0$_has_t       -- $\cmt{e_0 = e}$
     AddStep e$_0$ e$_1$ e$_0$_step_e$_1$ e e$_1$_eval_e ->  -- $\cmt{e_0 \step \evalsTo{e_1}{e}}$
       soundness e$_1$ t e (preservation e$_0$ t e$_0$_has_t e$_1$ e$_0$_step_e$_1$) e$_1$_eval_e
\end{mcode}
The reflexive case is proved by \ha{progress}.
In the inductive case the evaluation
sequence is $e_0 \step \evalsTo{e_1}{e}$
and the proof goes by induction,
using preservation to ensure that
$e_1$ is typed.

\subsection{Mechanization Details}

We provide a full, mechanically checked
proof of the metatheory in \S~\ref{sec:soundnessF}
and \S~\ref{sec:soundness}.
The only facts that are assumed are the
Requirements \ref{lem:prim-typing} and \ref{lem:implication}, 
resp. on built-in primitives
and the
implication relation.


\mypara{Representing Binders}
In our mechanization, we use the
\emph{locally-nameless representation} \cite{AydemirCPPW08,Chargueraud12}.
Free variables and bound variables
are taken to be separate syntactic
objects, so we do not need to worry
about alpha renaming of free variables
to avoid capture in substitutions.
We also use de Bruijn indices only
for bound variables. This enables us to avoid
taking binder names into account in the
$\mathsf{strengthen}$ function used to define
substitution (\cref{fig:type-subst}).
%

\mypara{Quantitative Results}
In~\cref{fig:empirical} we give
the empirical details of our metatheory,
which was checked using \lh version 0.8.10.7.1
and a Lenovo ThinkPad T15p laptop with
an Intel Core i7-11800H processor
with 8 physical cores and 32 GB of RAM.
Our mechanized proof is substantial,
spanning about 9400 lines distributed
over about 35 files.
Currently, the whole proof can be checked
in about 30 minutes, which can make interactive
development difficult.
While incremental modular checking provides
a modicum of interactivity, improving the
ergonomics, \ie verification time and providing
actionable error messages, remains an important
direction for future work.

\begin{table}[t]
\setlength{\tabcolsep}{12pt}
\begin{tabular}{ lrrrr  }
  \toprule
  \textbf{Subject} & \textbf{Files} & \textbf{Spec. (LOC)} & \textbf{Proof (LOC)} & \textbf{Time (mins)} \\
  \hline
  Definitions             & 6      & 1805  &  374  &  1\\
  Basic Properties        & 8      &  646  & 2117  &  4\\
  $\sysf$ Soundness       & 4      &  138  &  685  &  3\\
  Weakening               & 4      &  379  &  467  &  1\\
  Substitution            & 4      &  458  &  846  &  7\\
  Exact Typing            & 2      &   70  &  230  &  4\\
  Narrowing               & 1      &   88  &  166  &  1\\
  Inversion of Typing     & 1      &  124  &  206  &  1\\
  Primitives              & 3      &  120  &  277  &  4\\
  $\sysrf$ Soundness      & 1      &   14  &  181  &  1\\
  \midrule
  \textbf{Total}          & 35     & 3842  & 5549  & 29\\
  \bottomrule
\end{tabular}
\caption{Empirical details of our mechanization. We partition the development into sets of
         modules pertaining to different region of \cref{fig:graph}, and for each region
         separate the lines of specification (\eg definitions and lemma statements)
         from those needed for proofs.}
\label{fig:empirical}
\end{table}

\section{Related Work}
\label{sec:related}

We discuss the most closely related work
on the meta-theory of unrefined and refined
type systems.

\mypara{Representing Binders}
Our development for \sysf (\S~\ref{sec:soundnessF})
follows the standard presentation of System F's
metatheory by \citet{TAPL}.
The main difference between the two metatheories
is that ours includes well-formedness of types and
environments, which help with mechanization \cite{Remy21}
and are crucial when formalizing refinements.
One of the main challenges in mechanization
of metatheories is the syntactic representation
of variables and binders~\cite{Aydemir05}.
The \emph{named} representation
has severe difficulties because
of variable capturing substitutions
and the \emph{nameless} (\aka de Bruijn)
requires heavy index shifting.
The variable representation
of $\sysrf$ is
\emph{locally nameless representation}~\cite{Pollack93,AydemirCPPW08},
that is, free variables are named, but
the bound variables are represented by
syntactically distinct deBruijn indices.
%
%
We chose this representation because
it clearly addresses
the following two problems with named bound variables
but nevertheless our
metatheory still resembles
the paper and pencil proofs (that we performed
before  mechanization):
First, when different refinements are strengthened
(as in \cref{fig:type-subst}) the variable
capturing problem reappears
because we are substituting underneath a binder.
Second, subtyping usually permits
alpha-renaming of variables,
which breaks a required invariant
that each \sysrf derivation tree
is a valid \sysf tree after erasure.

%
%


\mypara{Hybrid \& Contract Systems}
\citet{flanagan06} formalizes
a monomorphic lambda calculus with
refinement types that differs from our \sysrf
in three ways.
%
First,
the denotational soundness methodology
of~\citet{flanagan06}
connects subtyping with expression
evaluation.
We could not follow this approach
because encoding type denotations as
a data proposition requires a
negative occurrence (\S~\ref{sec:typing:implication}).
%
%
Second, in \cite{flanagan06} type checking
is hybrid: the developed system
is undecidable and inserts runtime
casts when subtyping cannot
be statically decided.
Third, the original system lacks polymorphism.
\citet{SekiyamaIG17} extended hybrid types
with polymorphism, but unlike \sysrf,
their system does not support semantic subtyping.
For example, consider a divide by zero-error.
The refined types for \ha{div} and 0 could
be given by
$\mathsf{div} :: \funcftype{\tint}{\breft{\tint}{n}{n \neq 0} \rightarrow {\tint}}$
and $0 :: \breft{\tint}{n}{n=0}$.
This system will compile \ha{div 1 0}
by inserting a cast on 0:
$\langle\breft{\tint}{n}{n=0} \Rightarrow \breft{\tint}{n}{n\neq 0}\rangle$,
causing a definite runtime failure
that could have easily been prevented
statically.
Having removed semantic subtyping,
the metatheory of~\cite{SekiyamaIG17}
is highly simplified.
%
%
Static refinement type systems
(as summarized by~\citet{sprite})
usually restrict the definition
of predicates to quantifier-free
first-order formulae that can be
\emph{decided} by SMT solvers.
This restriction is not preserved
by evaluation that can substitute
variables with any value, thus
allowing expressions that cannot
be encoded in decidable logics, like
lambdas, to seep into the predicates of types.
In contrast, we allow predicates to be
any language term (including lambdas)
to prove soundness via preservation
and progress: our meta-theoretical
results trivially apply to systems that,
for efficiency of implementation, restrict
their source languages.
Finally, none of the above systems (hybrid,
contracts or static refinement types) come
with a machine checked soundness proof.

\mypara{Refinement Types in Coq}
Our soundness formalization follows
the axiomatized implication relation
of \citet{LehmannTanter} that decides
subtyping (our rule \sBase) by without
formally connecting implication and
expression evaluation.
\citet{LehmannTanter}'s Coq formalization
of a monomorphic lambda calculus with
refinement types differs from \sysrf
in two ways.
First, their axiomatized implication
allows them to arbitrarily
restrict the language of refinements.
We allow refinements to be arbitrary
program terms and intend, in the future,
to connect our axioms to SMT solvers
or other oracles.
Second, \sysrf includes language
features like polymorphism,
existentials, and selfification
which are critical for path- and
context-sensitive refinement typing,
but make the metatheory more challenging.
%
%
\citet{kuncak-stainless} present System FR,
a polymorphic, refined language
with a mechanized metatheory
of about 30K lines of Coq.
Compared to our system, their notion
of subtyping is not semantic, but relies on
a reducibility relation.
For example, even though System FR
will deduce that \tpos is a subtype of
\tint, it will fail to derive that
$\tint \rightarrow \tpos$ is subtype
of  $\tpos \rightarrow \tint$
as reduction-based subtyping
cannot reason about contra-variance.
Because of this more restrictive notion
of subtyping, their mechanization
does not require either the indirection of
denotational soundness or the use of
an implication proving oracle.
Further, System FR's support
for polymorphism is limited
in that it disallows refinements
on type variables, thereby
precluding many practically
useful specifications. 

\mypara{Metatheory in \lh}
\texttt{LWeb}~\cite{lweb} also used \lh 
to prove metatheory, concretely 
non-interference of $\lambda_{\text{LWeb}}$,
a core calculus that extends the LIO formalism with database access. 
The \texttt{LWeb} proof did not use refined data propositions, 
which were not present at the time of its development, and 
thus it has two major weaknesses compared to our 
present development. 
First, \texttt{LWeb}
\textit{assumes} termination of $\lambda_{\text{LWeb}}$'s evaluation function; 
without refined data propositions metatheory can be developed only over 
terminating functions. 
This was not a critical limitation since non-interference was only 
proved for terminating programs. 
However, in our proof the requirement that evaluation of \sysrf terminates 
would be too strict. 
In our encoding with refined data propositions
such an assumption was not required. 
Second, the \texttt{LWeb} development is not constructive: 
the structure of an assumed evaluation tree is logically inspected 
instead of the more natural case splitting permitted only with 
refined data propositions. 
This constructive way to develop metatheories is more compact 
(\eg there is no need to logically inspect the structure 
of the derivation trees) and more akin to the standard 
meta-theoretic developments of constructive tools like 
\coq and \isabelle. 
\section{Conclusions \& Future Work}

We presented and formalized soundness of
\sysrf, a core refinement calculus
that combines semantic subtyping,
existential types, and parametric
polymorphism, which are critical for
practical refinement typing but whose
combination has never been formalized.
Our meta-theory is mechanized in \lh,
using the novel feature of refined
data propositions to reify \eg typing
and evaluation derivations as (refined)
Haskell datatypes, and exploits SMT
to automate various tedious invariants
about variables.
We fully expect our proof can be
mechanized in specialized proof
systems like \agda \cite{agda},
\coq \cite{coq-book} or \isabelle \cite{NPW2002}
or those equipped with SMT-based
automation like \dafny \cite{Dafny},
or \fstar \cite{metafstar}.
Our goal here is not to compare
against such systems.
Instead, our primary contribution is to,
for the first time, \emph{establish the soundness}
of the combination of features critical for practical
refinement typing, and secondarily, show that such
a proof can be \emph{mechanized as a plain program}
with refinement types.
Looking ahead, we envision two lines
of work on mechanizing metatheory \emph{of}
and \emph{with} refinement types.

\mypara{1. Mechanization of Refinements}
\sysrf covers a crucial but small
fragment of the features of modern refinement
type checkers. It would be interesting to extend
the language to include features like refined
datatypes, and abstract and bounded refinements.
Similarly, our current work axiomatizes the
requirements of the semantic implication
checker (\ie SMT solver).
It would be interesting to implement a solver
and verify that it satisfies that contract, or
alternatively, show how proof certificates \cite{pcc}
could be used in place of such axioms.

\mypara{2. Mechanization with Refinements}
While this work shows that non-trivial
meta-theoretic proofs are \emph{possible}
with SMT-based refinement types, our experience
is that much remains to make such developments
\emph{pleasant}.
For example, programming would be far more convenient
with support for automatically \emph{splitting cases}
or filling in \emph{holes} as done in Agda \cite{agda}
and envisioned by \citet{hole_driven_liquid}.
Similarly, when a proof fails, the user has little
choice but to think really hard about the internal
proof state and what extra lemmas are needed to prove
their goal.
Finally, the stately pace of verification --- 9400 lines
across 35 files take about 30 minutes --- hinders
interactive development.
Thus, rapid incremental checking, lightweight synthesis,
and actionable error messages would go a long way towards
improving the ergonomics of verification, and hence remain
important directions for future work.


\bibliography{sw}

\appendix
\section{Proofs of \sysf Soundness}
\label{sec:proofsF}
In this appendix, we present the proofs for 
each lemma of our \sysf metatheory presented in \S~\ref{sec:soundnessF}.
\begin{theorem} (Type Safety)
    If $\hasftype{\varnothing}{\sexpr}{\sftype}$ and $\evalsTo{\sexpr}{\sexpr'}$,
    then $\sexpr'$ is a value or $\sexpr' \step \sexpr''$ 
    for some $\sexpr''$. 
\end{theorem}
\begin{proof}
We proceed by induction on the number of steps in $\evalsTo{\sexpr}{\sexpr'}$.
There are two cases for $\evalsTo{\sexpr}{\sexpr'}$: either $\sexpr=\sexpr'$ 
or there exists a term $\sexpr_1$ 
such that $\sexpr \step \evalsTo{\sexpr_1}{\sexpr'}$.
In the former case we conclude immediately by the Progress Lemma.
In the latter case, $\hasftype{\varnothing}{\sexpr_1}{\sftype}$
by the Preservation Lemma. Then by the inductive hypothesis
applied to $\sexpr_1$, we conclude that either $\sexpr'$ is a value 
or $\sexpr' \step \sexpr''$ for some $\sexpr''$. 
\end{proof}
\subsection{Progress}
\begin{lemma} (Progress) \label{lem:progressF-a} 
    If $\hasftype{\varnothing}{\sexpr}{\sftype}$ 
    then $\sexpr$ is a value or $\sexpr \step \sexpr'$ for some $\sexpr'$.
\end{lemma}
\begin{proof} 
We proceed by induction of the structure of 
$\hasftype{\varnothing}{\sexpr}{\sftype}$. In the cases of rule
\fPrim, \fVar, \fAbs, or \fTAbs, $\sexpr$ is a value.
\begin{itemize}
\pfcase{\fApp}: We have 
$\hasftype{\varnothing}{\sexpr}{\sftype}$ where
$\sexpr \equiv \app{\sexpr_1}{\sexpr_2}$. 
Inverting, we have that there exists some type $\sftype_2$
such that $\hasftype{\varnothing}{\sexpr_1}{\funcftype{\sftype_2}{\sftype}}$
and $\hasftype{\varnothing}{\sexpr_2}{\sftype_2}$.
We split on five possible cases for 
the structure of $\sexpr_1$ and $\sexpr_2$. 
First, suppose $\sexpr_1 \equiv \vabs{x}{\sexpr_0}$ and $\sexpr_2$ is 
a value. Then by rule \eAppAbs, 
$e \equiv \app{\vabs{x}{\sexpr_0}}{\sexpr_2} \step \subst{\sexpr_0}{x}{\sexpr_2}$.
Second, suppose $\sexpr_1 \equiv \vabs{x}{\sexpr_0}$ and $\sexpr_2$
is not a value. Then by the inductive hypothesis, there exists a term
$\sexpr'_2$ such that $\sexpr_2 \step \sexpr'_2$. Then by rule \eAppV
$e \equiv \app{\vabs{x}{\sexpr_0}}{\sexpr_2} \step \app{\vabs{x}{\sexpr_0}}{\sexpr'_2}$.
Third, suppose $\sexpr_1 \equiv \sconst$, a built in primitive 
and $\sexpr_2$ is a value. Then by rule \ePrim, 
$e \equiv \app{\sconst}{\sexpr_2} \step \delta(\sconst,\sexpr_2)$,
which is well-defined by the primitives lemma.
Fourth, suppose $\sexpr_1 \equiv \sconst$ and $\sexpr_2$
is not a value. Then by the inductive hypothesis, there exists a term
$\sexpr'_2$ such that $\sexpr_2 \step \sexpr'_2$. Then by rule \eAppV
$e \equiv \app{\sconst}{\sexpr_2} \step \app{\sconst}{\sexpr'_2}$.
Finally, by the canonical forms lemma, $\sexpr_1$ cannot be any other
value, so it must not be a value. Then by the inductive hypothesis,
there is a term $\sexpr'_1$ such that $\sexpr_1 \step \sexpr'_1$. 
Then by rule \eApp,
$e \equiv \app{\sexpr_1}{\sexpr_2} \step \app{\sexpr'_1}{\sexpr_2}$.
\pfcase{\fTApp}: We have 
$\hasftype{\varnothing}{\sexpr}{\sftype}$ where
$\sexpr \equiv \tyapp{\sexpr_1}{\stype}$ and
$\sftype \equiv \subst{\sigma}{\tvar}{\forgetreft{\stype}}$. 
Inverting, we have that 
$\hasftype{\varnothing}{\sexpr_1}{\polytype{\tvar}{\skind}{\sigma}}$.
We split on three cases for the structure of $\sexpr_1$.
First, suppose $\sexpr_1 \equiv \tabs{\tvar'}{\skind'}{\sexpr_0}$.
Then by rule \eTAppAbs, 
$\sexpr \equiv \tyapp{\tabs{\tvar'}{\skind'}{\sexpr_0}}{\stype} 
\step \subst{\sexpr_0}{\tvar'}{\stype}$.
Second, suppose $\sexpr_1 \equiv \sconst$, a built in primitive.
Then by rule \eTPrim, 
$\sexpr \equiv  \tyapp{\sconst}{\stype} \step \delta_T(\sconst,\forgetreft{\stype})$,
which is well-defined by the primitives lemma.
Finally, by the canonical forms lemma, $sexpr_1$ cannot be any other
form of value, so it must not be a value. Then by the inductive hypothesis,
there is a term $\sexpr'_1$ such that $\sexpr_1 \step \sexpr'_1$. 
Then by rule \eTApp
$\sexpr \equiv \tyapp{\sexpr_1}{\stype} \step \tyapp{\sexpr'_1}{\stype}$.
\pfcase{\fLet}: We have 
$\hasftype{\varnothing}{\sexpr}{\sftype}$ where
$\sexpr \equiv \eletin{x}{\sexpr_1}{\sexpr_2}$. Inverting,
we have that $\hasftype{\varnothing}{\sexpr_1}{\sftype_1}$
for some type $\sftype_1$. 
By the inductive hypothesis, either $\sexpr_1$ is a value
or there is a term $\sexpr'_1$ such that $\sexpr_1 \step \sexpr'_1$.
In the former case, rule \eLetV gives us
$e \equiv \eletin{x}{\sexpr_1}{\sexpr_2} \step \subst{\sexpr_2}{x}{\sexpr_1}$.
In the latter case, by rule \eLet, 
$e \equiv \eletin{x}{\sexpr_1}{\sexpr_2} \step \eletin{x}{\sexpr'_1}{\sexpr_2}$.
\pfcase{\fAnn}: We have 
$\hasftype{\varnothing}{\sexpr}{\sftype}$ where
$\sexpr \equiv \tyann{\sexpr_1}{\stype}$. Inverting,
we have the $\hasftype{\varnothing}{\sexpr_1}{\sftype}$ and
$\sftype = \forgetreft{\stype}$. 
By the inductive hypothesis, either $\sexpr_1$ is a value
or there is a term $\sexpr'_1$ such that $\sexpr_1 \step \sexpr'_1$.
In the former case, by rule \eAnnV, 
$e \equiv \tyann{\sexpr_1}{\stype} \step \sexpr_1$.
In the latter case, rule \eAnn gives us 
$e \equiv \tyann{\sexpr_1}{\stype} step \tyann{\sexpr'_1}{\stype}$.
\end{itemize}
\end{proof}
The progress proof is substantially the same for $\sysrf$. 
The only difference is that there is 
another straightforward inductive case for rule \tSub. 
\begin{lemma}\label{lem:canonicalF-a}
    ($\sysf$ Canonical Forms) \begin{enumerate}
    \item If $\hasftype{\varnothing}{v}{\Bool}$ 
        then $v = {\tt true}$ or $v = {\tt false}$.
    \item If $\hasftype{\varnothing}{v}{\Int}$ 
        then $v$ is an integer constant.
    \item If $\hasftype{\varnothing}{v}{\funcftype{\sftype}{\sftype'}}$ 
        then either $v = \vabs{x}{\sexpr}$ or $v = \sconst$, 
        a built in primitive function where 
        $\sconst \in \{\wedge, \vee, \neg, \leftrightarrow,\leq,=\}$.
    \item If $\hasftype{\varnothing}{v}{\polytype{a}{k}{\tau}}$ 
        then either $v = \Lambda a\bindt k.\, e$ 
        or $v$ is the polymorphic equality $=$.
    \item If $\isWFFT{\varnothing}{\tau}{\skbase}$ 
        then $\tau = \tbool$ or $\tau = \tint$.
    \end{enumerate}
\end{lemma}
\begin{proof}
    Parts (1) - (4) are easily deduced from the \sysf typing rules 
    in Figure \ref{fig:t} and the definition of $ty(c)$. 
    Part (5) is clear from the well-formedness rules in Figure \ref{fig:wf}.
\end{proof}
Lemma \ref{lem:canonicalF-a} is sufficient for our $\sysrf$ metatheory.
Our syntactic typing judgments in $\sysrf$ respect those of $\sysf$.
Specifically, if $\hastype{\tcenv}{\sexpr}{\stype}$ and 
$\isWellFormedE{\tcenv}$, then
$\hasftype{\forgetreft{\tcenv}}{\sexpr}{\forgetreft{\stype}}$.
Therefore, we do not have to state and 
prove a separate Canonical Forms Lemma for $\sysrf$.
\begin{lemma}\label{lem:inversionF-a} (Inversion of Typing) 
    \begin{enumerate}
        \item If $\hasftype{\tcenv}{c}{\sftype}$ 
            then $\sftype = \forgetreft{\ty{c}}$.
        \item If $\hasftype{\tcenv}{x}{\sftype}$ 
            then $\bind{x}{\sftype} \in \tcenv$.
        \item If $\hasftype{\tcenv}{\app{e}{e'}}{\sftype}$
            then there is some type $\sftype_x$ such that  
            $\hasftype{\tcenv}{e}{\funcftype{\sftype_x}{\sftype}}$ and
            $\hasftype{\tcenv}{e'}{\sftype_x}$.
        \item If $\hasftype{\tcenv}{\vabs{x}{e}}{\sftype}$ 
            then $\sftype = \funcftype{\sftype_x}{\sftype'}$ and
            $\hasftype{\bind{y}{\sftype_x},\tcenv}{\subst{e}{x}{y}}{\sftype'}$
            for some $\notmem{y}{\tcenv}$ and well-formed $\sftype_x$.
        \item If $\hasftype{\tcenv}{\tyapp{e}{t}}{\sftype}$ then there is some
            type $\sigma$ and kind $\skind$ such that 
            $\hasftype{\tcenv}{e}{\polytype{\al}{\skind}{\sigma}}$
            and $\sftype = \subst{\sigma}{\al}{\forgetreft{t}}$.
        \item If $\hasftype{\tcenv}{\tabs{\al}{\skind}{e}}{\sftype}$ then
            there is some type $\sftype'$ and kind $\skind$ such that
            $\sftype = {\polytype{\al}{\skind}{\sftype'}}$ and
            $\hasftype{\bind{\al'}{\skind},\tcenv}{\subst{e}{\al}{\al'}}
            {\subst{\sftype'}{\al}{\al'}}$ for some $\notmem{\al'}{\tcenv}$.
        \item If $\hasftype{\tcenv}{\eletin{x}{e_x}{e}}{\sftype}$ then
            there is some type $\sftype_x$ and $\notmem{y}{\tcenv}$ such that
            $\hasftype{\tcenv}{e_x}{\sftype_x}$ and 
            $\hasftype{\bind{y}{\sftype_x},\tcenv}{\subst{e}{x}{y}}{\sftype}$.
        \item If $\hasftype{\tcenv}{\tyann{e}{t}}{\sftype}$ then 
            $\sftype = \forgetreft{t}$ and $\hasftype{\tcenv}{e}{\sftype}$.
    \end{enumerate}
\end{lemma}
\begin{proof}
This is clear from the definition of the typing rules for $\sysf$. Each premise
can match only one rule because the $\sysf$ rules are syntax directed.
\end{proof}
The Inversion of Typing Lemma does not hold in $\sysrf$ due to the subtyping
relation. For instance 
$\hastype{\bind{x}{\breft{\tint}{\vv}{\vv = 5}}}{x}{\tint}$ but
$\notmem{\bind{x}{\tint}}{\bind{x}{\breft{\tint}{\vv}{\vv = 5}}}$.
In Lemma \ref{lem:inversion-a} we state and prove an analogous result 
for $\sysrf$ in the two cases needed to prove progress and preservation.
\begin{lemma}\label{lem:primitivesF-a}(Primitives) 
For each built-in primitive $\sconst$, 
\begin{enumerate}
\item If $\forgetreft{\ty{\sconst}} = \funcftype{\sftype_x}{\sftype}$, 
        and $\hasftype{\varnothing}{\sval}{\sftype_x}$ 
        then 
        $\hasftype{\varnothing}{\delta(\sconst,\sval)}{\sftype}$.
\item If $\forgetreft{\ty{\sconst}} = \polytype{\al}{\skbase}{\sftype'}$, 
        and $\isWFFT{\varnothing}{\sftype}{\skbase}$, 
        then 
        $\hasftype{\varnothing}{\delta_T(\sconst,\sftype)}{\subst{\sftype'}{\al}{\sftype}}$.
\end{enumerate}
\end{lemma}
\begin{proof}
\begin{enumerate}
    \item First consider $\sconst \in \{\wedge, \vee, \leftrightarrow\}$. 
        Then $\forgetreft{\ty{\sconst}} = \funcftype{\tbool}{\funcftype{\tbool}{\tbool}}$.
        Then by Lemma \ref{lem:canonicalF-a}, $\hasftype{\varnothing}{\sval}{\tbool}$
        gives us that $\sval = \ttrue$ or $\sval = \tfalse$.
        For each possibility for $\sconst$ and $\sval$, we can build a judgment 
        that $\hasftype{\varnothing}{\delta(\sconst,\sval)}{\funcftype{\tbool}{\tbool}}$.
        Similarly, if $\sconst = \neg$ 
        then $\forgetreft{\ty{\sconst}} = \funcftype{\tbool}{\tbool}$ and 
        $\delta(\neg,\sval) \in \{\ttrue,\tfalse\}$ can be typed at $\tbool$.
        The analysis for the other monomorphic primitives is entirely similar.
    \item Here $\sconst$ is the polymorphic $=$ and 
        $\forgetreft{\ty{\sconst}} = \polytype{\tvar}{\skbase}{\funcftype{\al}{\funcftype{\al}{\tbool}}}$. By the Canonical Forms Lemma, 
        $\sftype = \tbool$ or $\sftype = \tint$. In the former case,
        $\delta_T(\sconst,\tbool) = \leftrightarrow$, which we can type at 
        $\funcftype{\tbool}{\funcftype{\tbool}{\tbool}} =\subst{\forgetreft{\ty{\sconst}}}{\tvar}{\tbool}$. The case of $\tint$ is entirely similar
        because $\delta_T(\sconst,\tint)$ is the monomorphic integer equality.
\end{enumerate}
\end{proof}
\subsection{Preservation}
\begin{lemma} (Preservation) \label{lem:preservationF} 
    If $\hasftype{\varnothing}{\sexpr}{\sftype}$ and $\sexpr \step \sexpr'$, 
    then $\hasftype{\varnothing}{\sexpr'}{\sftype}$.
\end{lemma}   

\begin{proof} 
    We proceed by induction of the structure of 
    $\hasftype{\varnothing}{\sexpr}{\sftype}$. The cases of rules
    \fPrim, \fVar, \fAbs, or \fTAbs cannot occur because $\sexpr$ is a value
    and no value can take a step in our semantics.
    \begin{itemize}
    \pfcase{\fApp}: We have 
    $\hasftype{\varnothing}{\sexpr}{\sftype}$ where
    $\sexpr \equiv \app{\sexpr_1}{\sexpr_2}$. 
    Inverting, we have that there exists some type $\sftype_2$
    such that $\hasftype{\varnothing}{\sexpr_1}{\funcftype{\sftype_2}{\sftype}}$
    and $\hasftype{\varnothing}{\sexpr_2}{\sftype_2}$.
    We split on five possible cases for 
    the structure of $\sexpr_1$ and $\sexpr_2$. 
    First, suppose $\sexpr_1 \equiv \vabs{x}{\sexpr_0}$ and $\sexpr_2$ is 
    a value. Then by rule \eAppAbs and the determinism of our semantics, 
    $e' \equiv \subst{\sexpr_0}{x}{\sexpr_2}$.
    By the Inversion of Typing, for some $y$ we have
    $\hasftype{\bind{y}{\sftype_2}}{\subst{\sexpr_0}{x}{y}}{\sftype}$.
    By the Substitution Lemma, substituting $\sexpr_2$ through for $y$
    gives us $\hasftype{\varnothing}{\subst{\sexpr_0}{x}{\sexpr_2}}{\sftype}$
    as desired because $\subst{\subst{\sexpr_0}{x}{y}}{y}{\sexpr_2} = \subst{\sexpr_0}{x}{\sexpr_2}$.
    Second, suppose $\sexpr_1 \equiv \vabs{x}{\sexpr_0}$ and $\sexpr_2$
    is not a value. Then by the progress lemma, there exists a term
    $\sexpr'_2$ such that $\sexpr_2 \step \sexpr'_2$. Then by rule \eAppV
    and the determinism of our semantics,
    $e' \equiv \app{\vabs{x}{\sexpr_0}}{\sexpr'_2}$. 
    Now, by the inductive hypothesis, $\hasftype{\varnothing}{\sexpr'_2}{\sftype_2}$.
    Applying rule \fApp, $\hasftype{\varnothing}{\app{\sexpr_1}{\sexpr'_2}}{\sftype}$
    as desired.
    Third, suppose $\sexpr_1 \equiv \sconst$, a built in primitive, 
    and $\sexpr_2$ is a value. Then by rule \ePrim
    and the determinism of the semantics, 
    $e' \equiv \delta(\sconst,\sexpr_2)$.
    By the primitives lemma, 
    $\hasftype{\varnothing}{\delta(\sconst,\sexpr_2)}{\sftype}$ as desired.
    Fourth, suppose $\sexpr_1 \equiv \sconst$ and $\sexpr_2$
    is not a value. Then we argue in the same manner as the second case.
    Finally, by the canonical forms lemma, $\sexpr_1$ cannot be any other
    value, so it must not be a value. Then by the progress lemma,
    there is a term $\sexpr'_1$ such that $\sexpr_1 \step \sexpr'_1$. 
    Then by rule \eApp and the determinism of the semantics,
    $e' \equiv \app{\sexpr'_1}{\sexpr_2}$. By the inductive hypothesis,
    $\hasftype{\varnothing}{\sexpr'_1}{\funcftype{\sftype_2}{\sftype}}$.
    Applying rule \fApp, $\hasftype{\varnothing}{\app{\sexpr'_1}{\sexpr_2}}{\sftype}$
    as desired.
    \pfcase{\fTApp}: We have 
    $\hasftype{\varnothing}{\sexpr}{\sftype}$ where
    $\sexpr \equiv \tyapp{\sexpr_1}{\stype}$ and
    $\sftype \equiv \subst{\sigma}{\tvar}{\forgetreft{\stype}}$. 
    Inverting, we have that 
    $\hasftype{\varnothing}{\sexpr_1}{\polytype{\tvar}{\skind}{\sigma}}$
    and $\isWFFT{\varnothing}{\forgetreft{\stype}}{\skind}$.
    We split on three cases for the structure of $\sexpr_1$.
    First, suppose $\sexpr_1 \equiv \tabs{\tvar}{\skind}{\sexpr_0}$.
    Then by rule \eTAppAbs and the determinism of the semantics, 
    $\sexpr' \equiv \subst{\sexpr_0}{\tvar}{\stype}$.
    By the inversion of typing, for some $\tvar'$, we have
    $\hasftype{\bind{\tvar'}{\skind}}{\subst{\sexpr_0}{\tvar}{\tvar'}}
    {\subst{\sigma}{\tvar}{\tvar'}}$.
    By the Substitution Lemma, substituting $\forgetreft{\stype}$ 
    through for $\tvar$  gives us 
    $\hasftype{\varnothing}{\subst{\sexpr_0}{\tvar}{\stype}}{\subst{\sigma}{\tvar}{\forgetreft{\stype}}}$ as desired.
    Second, suppose $\sexpr_1 \equiv \sconst$, a built in primitive.
    Then by rule \eTPrim and the determinism of the semantics, 
    $\sexpr' \delta_T(\sconst,\forgetreft{\stype})$. By the primitives lemma,
    $\hasftype{\varnothing}{\delta_T(\sconst,\forgetreft{\stype})}{\subst{\sigma}{\tvar}{\forgetreft{\stype}}}$.
    Finally, by the canonical forms lemma, $sexpr_1$ cannot be any other
    form of value, so it must not be a value. Then by the progress lemma,
    there is a term $\sexpr'_1$ such that $\sexpr_1 \step \sexpr'_1$. 
    Then by rule \eTApp and the deterministic semantics
    $\sexpr' \equiv \tyapp{\sexpr'_1}{\stype}$.  By the inductive hypothesis,
    $\hasftype{\varnothing}{\sexpr'_1}{\polytype{\tvar}{\skind}{\sigma}}$.
    Applying rule \fTApp, 
    $\hasftype{\varnothing}{\tyapp{\sexpr'_1}{\stype}}{\subst{\sigma}{\tvar}{\forgetreft{\stype}}}$
    as desired.
    \pfcase{\fLet}: We have 
    $\hasftype{\varnothing}{\sexpr}{\sftype}$ where
    $\sexpr \equiv \eletin{x}{\sexpr_1}{\sexpr_2}$. Inverting,
    we have that 
    $\hasftype{\bind{y}{\sftype_1}}{\subst{\sexpr_2}{x}{y}}{\sftype}$
    and $\hasftype{\varnothing}{\sexpr_1}{\sftype_1}$
    for some type $\sftype_1$. 
    By the progress lemma either $\sexpr_1$ is a value
    or there is a term $\sexpr'_1$ such that $\sexpr_1 \step \sexpr'_1$.
    In the former case, rule \eLetV and determinism give us
    $\sexpr' \equiv \subst{\sexpr_2}{x}{\sexpr_1}$.
    By the Substitution Lemma (substituting $\sexpr_1$ for $x$), 
    we have $\hasftype{\varnothing}{\subst{\sexpr_2}{x}{\sexpr_1}}{\sftype}$
    as desired because 
    $\subst{\sexpr_2}{x}{\sexpr_1} = \subst{\subst{\sexpr_2}{x}{y}}{y}{\sexpr_1}$.
    In the latter case, by rule \eLet and determinism give us, 
    $\sexpr' \equiv \eletin{x}{\sexpr'_1}{\sexpr_2}$.
    By the inductive hypothesis we have that 
    $\hasftype{\varnothing}{\sexpr'_1}{\sftype_1}$ and by rule \fLet 
    we have $\hasftype{\varnothing}{\eletin{x}{\sexpr'_1}{\sexpr_2}}{\sftype}$. 
    \pfcase{\fAnn}: We have 
    $\hasftype{\varnothing}{\sexpr}{\sftype}$ where
    $\sexpr \equiv \tyann{\sexpr_1}{\stype}$. Inverting,
    we have the $\hasftype{\varnothing}{\sexpr_1}{\sftype}$ and
    $\sftype = \forgetreft{\stype}$. 
    By the progress lemma, either $\sexpr_1$ is a value
    or there is a term $\sexpr'_1$ such that $\sexpr_1 \step \sexpr'_1$.
    In the former case, by rule \eAnnV and the determinism of the semantics, 
    $\sexpr' \equiv \sexpr_1$. Then we already have that
    $\hasftype{\varnothing}{\sexpr'}{\sftype}$
    In the latter case, rule \eAnn and determinism give us 
    $\sexpr' \equiv \tyann{\sexpr'_1}{\stype}$. By the inductive hypothesis
    we have that $\hasftype{\varnothing}{\sexpr'_1}{\sftype}$. By rule
    \fAnn we conclude $\hasftype{\varnothing}{\tyann{\sexpr'_1}{\stype}}{\sftype}$.
    \end{itemize}
\end{proof}

The proof of preservation for $\sysrf$ differs in two cases above. 
In \tApp and \tTApp, we must use the Inversion of Typing lemma (\ref{lem:inversion-a})
from $\sysrf$ because the presence of rule \tSub prevents us from 
inferring the last rule used to type a term or type abstraction.
Furthermore, in case \tApp the substitution lemma would give us that 
$\hastype{\varnothing}{\sexpr'}{\subst{\stype}{x}{\sval_x}}$ for 
some value $\sval_x$. However we need to show preservation of the
existential type $\existype{x}{\stype_x}{\stype}$. This is done by
using rule \sWitn to show that, in fact, 
$\isSubType{\varnothing}{\subst{\stype}{x}{\sval_x}}{\existype{x}{\stype_x}{\stype}}$.
\begin{lemma}(Substitution)\label{lem:substitutionF}
    If $\hasftype{\tcenv}{\sval_x}{\sftype_x}$ 
    and if $\isWFFT{\tcenv}{\forgetreft{\stype_{\al}}}{k_{\al}}$ then 
    \begin{enumerate}
    \item If\; $\hasftype{\tcenv', \bind{x}{\sftype_x}, \tcenv}{\sexpr}{\sftype}$
        and $\isWFFE{\tcenv', \bind{x}{\sftype_x}, \tcenv}$ then
        ${\hasftype{\tcenv', \tcenv}{\subst{\sexpr}{x}{\sval_x}}{\sftype}}$.
    \item If\; $\hasftype{\tcenv', \bind{\al}{k_{\al}}, \tcenv}{\sexpr}{\sftype}$
        and $\isWFFE{\tcenv', \bind{\al}{k_{\al}}, \tcenv}$ then
        ${\hasftype{\subst{\tcenv'}{\al}{\forgetreft{\stype_{\al}}}, \tcenv}
                   {\subst{\sexpr}{\al}{\stype_{\al}}}
                   {\subst{\sftype}{\al}{\forgetreft{\stype_{\al}}}}}$
    \end{enumerate}
\end{lemma} 
\begin{proof}
We give the proofs for part (2); part (1) is similar but slightly
simpler because term variables do not appear in types in $\sysf$
We proceed by induction on the derivation tree of the typing judgment
$\hasftype{\tcenv', \bind{\al}{k_{\al}}, \tcenv}{\sexpr}{\sftype}$.
\pfcase{\fPrim}: We have $\sexpr \equiv \sconst$ and
$\hasftype{\tcenv', \bind{\al}{k_{\al}}, \tcenv}{\sconst}{\forgetreft{\ty{\sconst}}}$.
Neither $\sconst$ nor $\ty{\sconst}$ has any free variables, so each is 
unchanged under substitution. Then by rule \tPrim we conclude
$\hasftype{\subst{\tcenv'}{\al}{\forgetreft{\stype_{\al}}}, \tcenv}{\sconst}{\forgetreft{\ty{\sconst}}}$ because the environment may be chosen arbitrarily.
\pfcase{\fVar}: We have $\sexpr \equiv x$; by inversion, we get that 
$\bind{x}{\sftype} \in \tcenv', \bind{\al}{k_{\al}}, \tcenv$. We must have
$\tvar \neq x$ so there are two cases to consider for where $x$ can 
appear in the environment. If $\bind{x}{\sftype} \in \tcenv$,  then $\sftype$
cannot contain $\tvar$ as a free variable because $x$ is bound first in the environment (which grows from right to left). Then
$\hasftype{\subst{\tcenv'}{\al}{\forgetreft{\stype_{\al}}}, \tcenv}{x}{\sftype}$
as desired because $\subst{\sftype}{\tvar}{\forgetreft{\stype_{\al}}} = \sftype$.
Otherwise $\bind{x}{\sftype} \in \tcenv'$ and so 
$\bind{x}{\subst{\sftype}{\tvar}{\forgetreft{\stype_{\al}}}} \in \subst{\tcenv'}{\al}{\forgetreft{\stype_{\al}}}, \tcenv$. Thus
$\hasftype{\subst{\tcenv'}{\al}{\forgetreft{\stype_{\al}}}, \tcenv}{x}{\subst{\sftype}{\tvar}{\forgetreft{\stype_{\al}}}}$.
(In part (1), we have an additional case where 
$\hasftype{\tcenv', \bind{x}{\sftype}, \tcenv}{x}{\sftype}$. We have
$\subst{x}{x}{\sval_x} = \sval_x$ and so we can apply the weakening Lemma
to $\hasftype{\tcenv}{\sval_x}{\sftype}$ to obtain
$\hasftype{\tcenv', \tcenv}{\sval_x}{\sftype}$.)
\pfcase{\fApp}: We have $\sexpr \equiv \app{\sexpr_1}{\sexpr_2}$. By inversion
we have that
$\hasftype{\tcenv', \bind{\al}{k_{\al}}, \tcenv}{\sexpr_1}{\funcftype{\sftype_x}{\sftype}}$
and $\hasftype{\tcenv', \bind{\al}{k_{\al}}, \tcenv}{\sexpr_2}{\sftype_x}$.
Applying the inductive hypothesis to both of these, we get
$\hasftype{\subst{\tcenv'}{\al}{\forgetreft{\stype_{\al}}}, \tcenv}{\subst{\sexpr_1}{\tvar}{\stype_{\tvar}}}{\subst{\funcftype{\sftype_x}{\sftype}}{\tvar}{\forgetreft{\stype_{\tvar}}}}$
and
$\hasftype{\subst{\tcenv'}{\al}{\forgetreft{\stype_{\al}}}, \tcenv}{\subst{\sexpr_2}{\tvar}{\stype_{\tvar}}}{\subst{\sftype_x}{\tvar}{\forgetreft{\stype_{\tvar}}}}$. 
Combining these by rule \fApp, we conclude
$\hasftype{\subst{\tcenv'}{\al}{\forgetreft{\stype_{\al}}}, \tcenv}{\subst{\app{\sexpr_1}{\sexpr_2}}{\tvar}{\stype_{\tvar}}}{\subst{\sftype}{\tvar}{\forgetreft{\stype_{\tvar}}}}$.
\pfcase{\fAbs}: We have $\sexpr \equiv \vabs{x}{\sexpr_1}$ 
and $\sftype \equiv \funcftype{\sftype_x}{\sftype_1}$. 
By inversion we have that for some $y$, both
$\hasftype{\bind{y}{\sftype_x},\tcenv', \bind{\al}{k_{\al}}, \tcenv}
{\subst{\sexpr_1}{x}{y}}{\sftype_1}$
and
$\isWFFT{\tcenv', \bind{\al}{k_{\al}}, \tcenv}{\sftype_x}{\skind_x}$.
By the inductive hypothesis, and the Substitution Lemma for 
well-formedness judgments, we have
$\hasftype{\bind{y}{\subst{\sftype_x}{\al}{\forgetreft{\stype_{\al}}}},\subst{\tcenv'}{\al}{\forgetreft{\stype_{\al}}}, \tcenv}
{\subst{\subst{\sexpr_1}{\al}{\stype_{\al}}}{x}{y}}{\subst{\sftype_1}{\al}{\forgetreft{\stype_{\al}}}}$
and
$\isWFFT{\subst{\tcenv'}{\al}{\forgetreft{\stype_{\al}}}, \tcenv}
{\subst{\sftype_x}{\al}{\forgetreft{\stype_{\al}}}}{\skind_x}$,
where we can switch the order of substitutions because $y$ does not 
appear free in the well-formed type $\stype_{\tvar}$.
Then we can conclude by applying rule \fAbs that 
$\hasftype{\subst{\tcenv'}{\al}{\forgetreft{\stype_{\al}}}, \tcenv}
{\subst{\vabs{x}{\sexpr_1}}{\tvar}{\stype_\tvar}}
{\subst{\funcftype{\sftype_x}{\sftype_1}}{\tvar}{\forgetreft{\stype_\tvar}}}$.
\pfcase{\fLet}: We have $\sexpr \equiv \eletin{x}{\sexpr_1}{\sexpr_2}$
and by inversion we have that for some type $\sftype_1$,
$\hasftype{\tcenv', \bind{\al}{k_{\al}}, \tcenv}{\sexpr_1}{\sftype_1}$
and for some $\notmem{y}{\tcenv', \bind{\al}{k_{\al}}, \tcenv}$,
$\hasftype{\bind{y}{\sftype_1},\tcenv', \bind{\al}{k_{\al}}, \tcenv}
{\subst{\sexpr_2}{x}{y}}{\sftype}$.
By the inductive hypothesis, we have that
$\hasftype{\subst{\tcenv'}{\al}{\forgetreft{\stype_{\al}}}, \tcenv}
{\subst{\sexpr_1}{\tvar}{\stype_\tvar}}{\subst{\sftype_1}{\tvar}{\forgetreft{\stype_\tvar}}}$
and
$\hasftype{\bind{y}{\subst{\sftype_1}{\tvar}{\stype_\tvar}},\subst{\tcenv'}{\al}{\forgetreft{\stype_{\al}}}, \tcenv}
{\subst{\sexpr_2}{x}{\stype_\tvar}}{\subst{\sftype}{\tvar}{\forgetreft{\stype_\tvar}}}$.
Then by rule \fLet we conclude
$\hasftype{\subst{\tcenv'}{\al}{\forgetreft{\stype_{\al}}}, \tcenv}
{\eletin{x}{\subst{\sexpr_1}{\tvar}{\stype_\tvar}}{\subst{\sexpr_2}{\tvar}{\stype_\tvar}}}{\subst{\sftype}{\tvar}{\forgetreft{\stype_\tvar}}}$.
\pfcase{\fAnn}: We have $\sexpr \equiv \tyann{\sexpr'}{\stype}$ 
and by inversion we have that $\forgetreft{\stype} = \sftype$ and 
$\hasftype{\tcenv', \bind{\al}{k_{\al}}, \tcenv}
{\sexpr'}{\sftype}$.
By the inductive hypothesis, we have
$\hasftype{\subst{\tcenv'}{\al}{\forgetreft{\stype_{\al}}}, \tcenv}
{\subst{\sexpr'}{\tvar}{\stype_\tvar}}{\subst{\sftype}{\tvar}{\forgetreft{\stype_\tvar}}}$.
By our definition of refinement erasure, we have
$\forgetreft{\subst{\stype}{\tvar}{\stype_\tvar}}
 = \subst{\forgetreft{\stype}}{\tvar}{\forgetreft{\stype_\tvar}}$
and we have 
$\subst{\tyann{\sexpr'}{\stype}}{\tvar}{\stype_\tvar} 
 = \tyann{\subst{\sexpr'}{\tvar}{\stype_\tvar}}{\subst{\stype}{\tvar}{\stype_\tvar}}$. Thus by rule \fAnn,
$\hasftype{\subst{\tcenv'}{\al}{\forgetreft{\stype_{\al}}}, \tcenv}
{\subst{\tyann{\sexpr'}{\stype}}{\tvar}{\stype_\tvar}}{\subst{\forgetreft{\stype}}{\tvar}{\forgetreft{\stype_\tvar}}}$.
\end{proof}

\section{Proofs of System RF Soundness}
\label{sec:proofs}

We present the proofs in this appendix in the same order 

\mypara{Inversion of Typing Judgments.} 
In the yellow region of Figure \ref{fig:graph}, we discuss how the fact that the typing judgment is no longer syntax-directed leads to an involved proof for the inversion lemma below. First, we need a lemma about subtyping:

\begin{lemma}
    {\em Transitivity of Subtyping}: If $\isWellFormed{\tcenv}{t}{k}$, $\isWellFormed{\tcenv}{t'}{k'}$, $\isWellFormed{\tcenv}{t''}{k''}$, and $\vdash_w \tcenv$ and $\isSubType{\tcenv}{t}{t'}$ and $\isSubType{\tcenv}{t'}{t''}$ then $\isSubType{\tcenv}{t}{t''}$. (LemmasTransitive.hs) Our ${\tt lem\_sub\_trans}$ depends on ${\tt lem\_subst\_sub}$.
\end{lemma}

\begin{proof}
    We proceed by induction on the combined size of the derivation trees of $\tcenv \vdash t' <: t'$ and $\tcenv \vdash t' <: t''$.
    We first consider the cases where none of $t, t',$ or $t''$ are existential types. By examination of the subtyping rules, we see that $t$, $t'$, and $t''$ must have the same form, and so there are three such cases:
    
    \pfcase{Refinement types}: 
    In this case $t \equiv b\{x_1\col p_1\}$, $t' \equiv b\{x_2\col p_2\}$, and $t'' \equiv b\{x_3\col p_3\}$. The last rule used in each of $\tcenv \vdash t' <: t'$ and $\tcenv \vdash t' <: t''$ must have been {\sc S-Base}. Inverting each of these we have
    \begin{equation}
    \entails{y\bindt b\{x_1\col p_1\}, \tcenv}{p_2[y/x_2]} \;\;\;\;{\rm and}\;\;\;\; \entails{z\bindt b\{x_2\col p_2\}, \tcenv}{p_3[z/x_3]}
    \end{equation}
    for some $y,z \not\in\dom{\tcenv}$. We can invert {\sc Ent-Pred}  to get
    \begin{equation}
    \forall \theta.\, \theta \in\lb y\bindt b\{x_1\col p_1\}, \tcenv\rb \Rightarrow \theta(p_2[y/x_2]) \steps \ttrue
    \end{equation}and\begin{equation}
    \forall \theta.\, \theta \in\lb y\bindt b\{x_2\col p_2\}, \tcenv\rb \Rightarrow \theta(p_3[y/x_3]) \steps \ttrue.
    \end{equation}
    where we also use the change of free variables lemma in the last equation.
    Let $\theta \in\lb y\bindt b\{x_1\col p_1\}, \tcenv\rb$. 
    Let $v = \theta(y)$ Then $v \in \lb \theta(b\{x_1\col p_1\})\rb,$ and so $v \in \lb \theta(b\{x_2\col p_2\})\rb$ by the Denotational Soundness Lemma. Then $\theta \in\lb y\bindt b\{x_2\col p_2\}, \tcenv\rb$ also, and so 
    $\theta(p_3)[y/x_3] \steps \ttrue.$
    Therefore we can conclude that
    \begin{equation}
        \entails{y\bindt b\{x_1\col p_1\}, \tcenv}{p_3[y/x_3]}
    \end{equation} 
    and thus that $\isSubType{\tcenv}{b\{x_1\col p_1\}}{b\{x_3\col p_3\}}$.
    
    \mypara{Case function types}: 
    In this case $t \equiv \functype{x_1}{s_1}{t_1}$, $t' \equiv \functype{x_2}{s_2}{t_2}$, and $t'' \equiv \functype{x_3}{s_3}{t_3}$. The last rule used in each of $\tcenv \vdash t' <: t'$ and $\tcenv \vdash t' <: t''$ must have been {\sc S-Func}. Inverting each of these we have
    \begin{equation*}
    \isSubType{\tcenv}{s_2}{s_1}, \;\;\; \isSubType{\tcenv}{s_3}{s_2}, \;\;\;
    \isSubType{y\bindt s_2, \tcenv}{t_1[y/x_1]}{t_2[y/x_2]}, \;\;{\rm and} \;\;
    \isSubType{z\bindt s_3, \tcenv}{t_2[z/x_2]}{t_3[z/x_3]}
    \end{equation*}
    for some $y,z \not\in\dom{\tcenv}$. By the inductive hypothesis we can combine the first two judgments to get $\isSubType{\tcenv}{s_3}{s1}$. By the narrowing lemma (see below) and the change of variables lemma we have $\isSubType{z\bindt s_3, \tcenv}{t_1[z/x_1]}{t_2[z/x_2]}$; then we conclude by the inductive hypothesis that $\isSubType{z\bindt s_3, \tcenv}{t_1[z/x_1]}{t_3[z/x_3]}$. 
    Then by rule {\sc S-Func} we conclude that $\isSubType{\tcenv}{\functype{x_1}{s_1}{t_1}}{\functype{x_3}{s_3}{t_3}}$.
    The well-formedness judgments required to apply the inductive hypothesis can be constructed by inverting $\isWellFormed{\tcenv}{t}{*}$, $\isWellFormed{\tcenv}{t'}{*}$, and $\isWellFormed{\tcenv}{t''}{*}$ and using the change of free variables and narrowing lemmas (for well-formedness).
    
    \mypara{Case polymorphic types}: In this case $t \equiv \polytype{\al_1}{k_1}{t_1}$, $t' \equiv \polytype{\al_2}{k_1}{t_2}$, and $t'' \equiv \polytype{\al_3}{k_1}{t_3}$. The last rule used in each of $\tcenv \vdash t' <: t'$ and $\tcenv \vdash t' <: t''$ must have been {\sc S-Poly}. Inverting each of these we have
    \begin{equation*}
    \isSubType{\al\bindt k_1, \tcenv}{t_1[\al/\al_1]}{t_2[\al/\al_2]}, \;\;\;{\rm and} \;\;\;
    \isSubType{\al'\bindt k_1, \tcenv}{t_2[\al'/\al_2]}{t_3[\al'/\al_3]}
    \end{equation*}
    for some $\al,\al' \not\in\dom{\tcenv}$. By the change of variables lemma, 
    we have $\isSubType{\al\bindt k_1, \tcenv}{t_2[\al/\al_2]}{t_3[\al/\al_3]}$. By the inductive hypothesis we can conclude that
    $\isSubType{\al\bindt k_1, \tcenv}{t_1[\al/\al_1]}{t_3[\al/\al_3]}$
    and thus
    $\isSubType{\tcenv}{\polytype{\al_1}{k_1}{t_1}}{\polytype{\al_3}{k_1}{t_3}}$.
    The well-formedness judgments required to apply the inductive hypothesis can be constructed by inverting $\isWellFormed{\tcenv}{t}{*}$, $\isWellFormed{\tcenv}{t'}{*}$, and $\isWellFormed{\tcenv}{t''}{*}$ and using the change of free variables lemma.

    \mypara{Case $t''$ }

    \mypara{}

    \mypara{}
    
    \end{proof}

\begin{lemma}
    {\em Inversion of TAbs/TAbsT}: If $\hastype{\tcenv}{(\lambda w. e)}{x\bindt t_x->t}$ and $\vdash_w \tcenv$ then there exists $y \not\in \dom(\tcenv)$ such that $\hastype{y\bindt tx,\tcenv}{e[y/w]}{t[y/x]}$.  If $\hastype{\tcenv}{(\Lambda a_1\bindt k_1. e)}{\forall a\bindt k. t}$ and $\vdash_w \tcenv$ then exists $a' \not\in \dom(\tcenv)$ such that $\hastype{a'\bindt k,\tcenv}{e[a'/a_1]}{t[a'/a]}$. 
\end{lemma}

\end{document}